\documentclass[a4paper,fleqn,usenatbib]{mnras}
\usepackage{newtxtext,newtxmath}
\usepackage[T1]{fontenc}
\usepackage{ae,aecompl}
\usepackage{graphicx}	
\usepackage{amsmath}	
\usepackage{amssymb}	
\usepackage{deluxetable}

\def\gtorder{\mathrel{\raise.3ex\hbox{$>$}\mkern-14mu
             \lower0.6ex\hbox{$\sim$}}}
\def\ltorder{\mathrel{\raise.3ex\hbox{$<$}\mkern-14mu
             \lower0.6ex\hbox{$\sim$}}}

\title[Constraints on the ejecta of NS mergers]{Constraints on the ejecta of the GW170817 neutron-star merger from its electromagnetic emission}
\author[E. Waxman et al.]{Eli Waxman, Eran O. Ofek, Doron Kushnir, Avishay Gal-Yam \\
Dept. of Particle Phys. \& Astrophys., Weizmann Institute of Science, Rehovot 76100, Israel}
\date{Accepted XXX. Received YYY; in original form ZZZ}
\pubyear{2015}

\begin{document}
\label{firstpage}
\pagerange{\pageref{firstpage}--\pageref{lastpage}}
\maketitle
\begin{abstract}
We present a simple analytic model, that captures the key features of the emission of radiation from material ejected by the merger of neutron stars (NS), and construct the multi-band and bolometric luminosity light curves of the transient associated with GW170817, AT\,2017gfo, using all available data. The UV to IR emission is shown to be consistent with a single $\approx0.05$\,M$_\odot$ component ejecta, with a power-law velocity distribution between $\approx 0.1\,c$ and $>0.3\,c$, a low opacity, {$\kappa<1$\,cm$^2$\,g$^{-1}$}, and a radioactive energy release rate consistent with an initial $Y_{\rm e}<0.4$.
The late time spectra require an opacity of $\kappa_\nu\approx0.1$\,cm$^2$\,g$^{-1}$ at 1 to $2\mu$m. If this opacity is provided entirely by Lanthanides, their implied mass fraction is $X_{\rm Ln}\approx10^{-3}$, approximately 30 times below the value required to account for the solar abundance. The inferred value of $X_{\rm Ln}$ is uncertain due to uncertainties in the estimates of IR opacities of heavy elements, which also do not allow the exclusion of a significant contribution to the opacity by other elements (the existence of a slower ejecta rich in Lanthanides, that does not contribute significantly to the luminosity, can also not be ruled out). The existence of a relatively massive, $\approx 0.05$\,M$_\odot$, ejecta with high velocity and low opacity is in tension with the results of numerical simulations of NS mergers.
\end{abstract}

\section{Introduction}
\label{sec:Intro}

The merger of a neutron star (NS) with its binary NS or black hole (BH) companion has been suggested to produce high density neutron rich ejecta, in which heavy elements beyond Iron are produced by the r-process \citep{LattimerSchramm74}. A strong optical-UV emission was predicted to be produced by the expanding ejecta, due to its heating by radioactive decay of unstable isotopes \citep{LiPac98}.

Several mechanisms are generally considered for the ejection of material: "Dynamical" ejection taking place on the merger time scale, $\sim1$~ms, due to tidal {interaction} or shock waves generated near the contact surface of the merging NSs; A "wind" driven by neutrinos emitted by a massive neutron star produced by the merger (which will be absent if the merger leads to a direct collapse to a BH); "Secular" ejection on longer time scale ($\lesssim$100~ms) from an accretion disk that is expected to form around the merger remnant \citep{Rosswog99,Oechslin07,MetzgerThompson08,Dessart09,MetzgerPiro09,FernandezMetzger13,Hotokezaka13,Perego14,JustBauswein15,MartinPerego15,Sekiguchi16,Wu16,Lippuner17,SiegelMetzger17,PeregoRadice173KN}. The dynamical ejecta is commonly expected to be fast, $v\sim0.3c$, and characterized by a low electron number fraction, $Y_{\rm e}<0.2$, leading to the formation of "heavy r-process" elements all the way up to atomic number $A\approx 200$. Theoretical estimates of the mass of this component of the ejecta range from $10^{-4}$ to $10^{-2}M_\odot$. The wind ejecta is commonly expected to be characterized by higher values of $Y_{\rm e}$, $Y_{\rm e}>0.2$, leading to the formation of "light r-process" elements, up to $A\approx 140$. Finally, the secular ejecta is expected to be relatively slow, $v\sim0.05\,c$, and massive, $\sim10^{-2}M_\odot$, and its characteristic $Y_{\rm e}$ value may range from 0.1 to 0.5, depending on the assumed life time of the massive neutron star produced by the merger.

Due to the large opacity of Lanthanide elements ($57\le Z\le 71$, $140\lesssim A\lesssim 180$), which leads to a long photon diffusion time through the ejecta, low $Y_{\rm e}$ components are expected to produce a dimmer, redder emission peaking on a few days time scale \citep[a "red kilo-nova",][]{Kasen13,BarnesKasen13}, while higher $Y_{\rm e}$ components are expected to produce a brighter, bluer emission on a day time scale (a "blue kilo-nova").

The recent discovery of gravitational wave emission from a NS binary merger, GW170817 \citep{Abbott17PhRvL}, followed by the detection of an electromagnetic counterpart, AT\,2017gfo \citep[e.g.][]{2017ApJ...848L..13A,2017arXiv171005452C,2017ApJ...848L..16S,2017ApJ...848L..24V}, opens the possibility for testing the theory of NS mergers. In particular, it enables one to use observations in order to determine the properties of the ejecta, including its velocity distribution, opacity, radioactivity and composition.

The electromagnetic emission observed following the NS binary merger is in general consistent with the emission of radiation from a mildly relativistic expanding ejecta, which is being continuously heated by radioactive energy release at a rate corresponding to radioactive elements heavier than the Iron group. Models have been constructed, that reasonably reproduce the observed radiation by summing the light curves and spectra of several (two or three) ejecta components \citep{Kasen17Nat,Rosswog17,CowperthwaiteBerger173KN,Tanaka17,PeregoRadice173KN}. A low-opacity, {$\kappa\approx 0.5{\rm cm^2/g}$,} "blue kilonova" component is required in these models to produce the early, $\lesssim2$~d, blue emission, and a high opacity, {$\kappa\approx 10{\rm cm^2/g}$,} "red-kilonova" component is required to produce the later, $t\gtrsim4$~d, redder emission \citep[models invoking contributions from additional jet/cocoon components have also been suggested, e.g.][]{2017arXiv171005436K,Piro17,Gottlieb18Cocoon}. \citet{2017Natur.551...75S} point out that the luminosity produced by the blue component may dominate the total luminosity also at late time, hence leaving little place for a red component contribution. They do not, however, analyze the color/spectra data available for $t>3$~d, and conclude that a high-opacity Lanthanide rich component may be required at late time.

The theoretical predictions for the electromagnetic emission are based on various combinations of detailed numerical calculations and semi-analytic approximations. For example: \citet{2017Natur.551...75S} and \citet{2017arXiv171011576V} use the semi-analytic model of \citet{Arnett82} for SN Ia light curves, with a modified radioactive energy release rate and a thermalization efficiency (of the radioactive decay products) based on the numerical calculations of \citet{Barneskasen16}; \citet{Rosswog17} and \citet{PeregoRadice173KN} use detailed hydrodynamic numerical simulations to describe the density and velocity fields of the ejecta, combined with a semi-analytic description of the deposition of energy and of the emission of radiation; \citet{Kasen17Nat} and \citet{Kilpatrick17} assume simple analytic spherical density and velocity distributions of the ejecta, and calculate the emission of radiation using a detailed numerical model of the energy deposition and radiation transfer. The two component models of \citet{Kasen17Nat} and \citet{Kilpatrick17}, for example, are based on such detailed numerical calculations of spherical kilonova shells over a grid of chosen shell masses, velocities and compositions (corresponding to a grid of initial values of $Y_{\rm e}$).

While the numerical modeling approach has some advantages, it also has a clear drawback. It relies on numerical simulations that depend on a considerable number of highly uncertain parameters. Exploring the dependence on a multiple number of highly uncertain parameters is difficult in the detailed numerical simulations approach, since significant computation time is required for obtaining the model results for each choice of the values of the parameters. This is particularly valid for the present case. As mentioned above, the ejecta mass, its velocity distribution and its $Y_{\rm e}$ distribution (and hence composition distribution) are all highly uncertain. Uncertainties exit also in both the radioactive heat production and deposition, the latter due largely to the uncertainties in the fractional energy carried by $\gamma$-rays, $\alpha$- and $\beta$-particles, and in the energy distribution of these secondary particles. Furthermore, the line structure and opacities of high-Z elements, that are relevant for this case, are not fully known. Specifically, the huge number of possible line transitions makes it difficult to establish trustable line lists, and calculate opacities. We note that even for iron, there are considerable uncertainties (see discussion in \S~\ref{sec:late}). Finally, we note that combining the contributions of several ejecta components by simply summing up their luminosities (and spectra) is not always valid, since the radiation emitted by one component may be modified as it passes through the others. Given these large uncertainties, we suggest a somewhat different approach.

We develop a simple analytic model that captures the key physics of kilo-nova (KN) emission (\S~\ref{sec:Model}). The model reproduces the key features obtained in detailed numerical calculations, and presents explicitly (and analytically) the dependence of the results on uncertain model parameters. It therefore enables one to infer the ejecta properties directly as functions of the parameters that describe the observations, as well as to estimate the sensitivity of the inferred parameters to uncertainties in both model parameters and observations. Furthermore, constraints on ejecta parameters, that are inferred using the simple model, may be used to guide more sophisticated numerical calculations.

In a nutshell, our model includes an expanding mass described by a power-law velocity distribution. {We assume that the ejecta is composed of a single component, in the sense that the microscopic plasma parameters (like composition and opacity) are assumed not to vary strongly across the ejecta}. We thus use a uniform radioactive energy release rate (per unit mass), $\dot{\varepsilon}(t)$, taking into account the spatial and temporal dependence of the fraction of the radioactive decay energy that is deposited in the plasma, and a uniform time dependent opacity, $\kappa\propto t^\gamma$. A uniform radioactive energy release is likely to be a good approximation {also in the case where the composition varies significantly across the ejecta}, since the energy release obtained for a wide range of $Y_{\rm e}<0.4$ is nearly independent of $Y_{\rm e}$ \citep[this approximation is more appropriate for the current case than the approximation of a centrally localized energy deposition, used by][for SN Ia modelling]{Arnett82}. A temporal evolution of the opacity is allowed in order to account for variations due to the temperature and density evolution of the plasma. Moreover, it may partially capture the effect of an opacity variation due to a composition gradient within the ejecta, since the opacity at the region from which photons are emitted would evolve with time in this case. We note, however, that this simple parametrization cannot capture large abrupt variations of the opacity, due to strong spatial or temporal evolution (see further discussion in \S~\ref{sec:ModelDesc}). Finally, we note that our model differs from those used in earlier calculations also in the allowed velocity distribution (we allow a wider velocity distribution), and in the description of the efficiency of radioactive energy deposition. This has a significant effect on the predicted light curves, as discussed in \S~\ref{sec:ModelDesc} (deposition efficiency) and \S~\ref{sec:comp_numerical} (velocity distribution).

The structure of the paper is as follows. In \S~\ref{sec:obs} we present the photometric and spectroscopic observations we used, as well as the construction of the multi-band light curves, the bolometric light curve, and the photometric and spectral evolution. The analytic model is presented in \S~\ref{sec:Model} (with a full derivation in the Appendix, \S~\ref{sec:ModelDer}). In \S~\ref{sec:comp_obs} we show that the simple model fully describes the IR to UV emission of the AT\,2017gfo GW electromagnetic counterpart, and infer the ejecta parameters. Finally, our conclusions are summarized and discussed in \S~\ref{sec:discussion}.

\section{Observations}
\label{sec:obs}

About 10\,hr following the GW event, an optical counterpart was identified \citep[e.g.][]{2017ApJ...848L..13A,2017arXiv171005452C,2017ApJ...848L..16S,2017ApJ...848L..24V}, that decayed on a time scale of about one week. Here we re-analyze the UV to IR photometric and spectroscopic observations of AT\,2017gfo. We do not discuss the $\gamma$-ray \citep{2017ApJ...848L..13A}, X-ray \citep{Haggard17,Troja17,Margutti17} and radio emission \citep{Hallinan17}, the origin of which is different from the radioactively powered IR-UV emission. In \S\ref{sec:bol} we describe the multi-band photometric and spectroscopic data we used, as well as the construction of the multi-band and bolometric light curves. In \S\ref{sec:fits} we fit a broken power-law and an exponential decay description for the bolometric light curve, and present our estimates of the effective temperature, photospheric radius and velocity as a function of time. Throughout, we assume a distance of 40\,Mpc to the merger. The analysis in this section was done using tools available\footnote{https://webhome.weizmann.ac.il/home/eofek/matlab/}
as part of the MATLAB Astronomy \& Astrophysics Toolbox \citep{2014ascl.soft07005O}.

\subsection{Multi-band and bolometric light curves}
\label{sec:bol}

The main source for the photometric data we used is \cite{2017arXiv171011576V} who collected the available data from
the literature
(\cite{2017Natur.551...64A};
\cite{2017arXiv171005452C};
\cite{2017ApJ...848L..17C};
\cite{2017ApJ...848L..29D};
\cite{2017arXiv171005443D};
\cite{2017arXiv171005437E};
\cite{2017arXiv171005462H};
\cite{2017arXiv171005436K};
\cite{2017ApJ...850L...1L};
\cite{2017Natur.551...67P};
\cite{2017arXiv171005448P};
\cite{Tanvir17};
\cite{2017Natur.551...75S};
\cite{2017Natur.551...71T};
\cite{2017arXiv171005848U};
\cite{2017ApJ...848L..24V};
\cite{2017arXiv171005432S}).
Several of these sources reduce the same data, and the reported magnitudes are not always consistent.
\cite{2017arXiv171011576V} attempted to correct some of these discrepancies.

In our analysis we started with the \cite{2017arXiv171011576V} multi-band photometric collection.
However, our analysis suggests that some photometric points still
suffer systematic offsets.
In order to address this problem, we plotted for each band
the light curve from all sources, and removed points with
large systematic errors.
We note that we removed all the photometric points by
\cite{2017Natur.551...71T},
as they show considerable systematic offsets from other sources.
We then fitted the light curve in each band with a polynomial
in $\log(t)$-magnitude space, were the degree of the polynomial (0 to 6)
was chosen subjectively, by eye.
The outcome (given in Table~\ref{tab:MultiBand} in \S~\ref{sec:data}) is that for each band we have a smooth polynomial
that describe the light curve between a start and end point
(which is $\approx0.1$ day before/after the first/last point).
The error in each light curve is taken as the rms of the polynomial fit.
Next, in a grid of predefined times, between $0.5$\,days
and $16.5$\,days after the merger, we read the magnitude of the transient,
in each band, from the interpolation polynomials.
\begin{figure}
\centerline{\includegraphics[width=8cm]{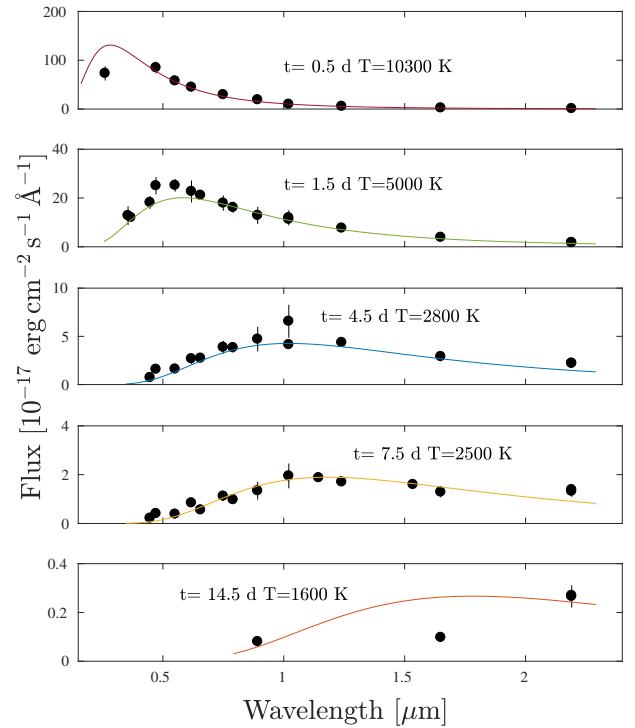}}
\caption{Examples for blackbody fits to the photometric data.
Black circles represent the photometric points interpolated using the interpolation
polynomials to some predefined times. Solid lines represents the best fit blackbody fit.
Time after the merger and best fit temperature are indicated on each plot. See also Table~\ref{tab:BolLum}.
\label{fig:PhotSpectBB}}
\end{figure}
\begin{figure}
\centerline{\includegraphics[width=8cm]{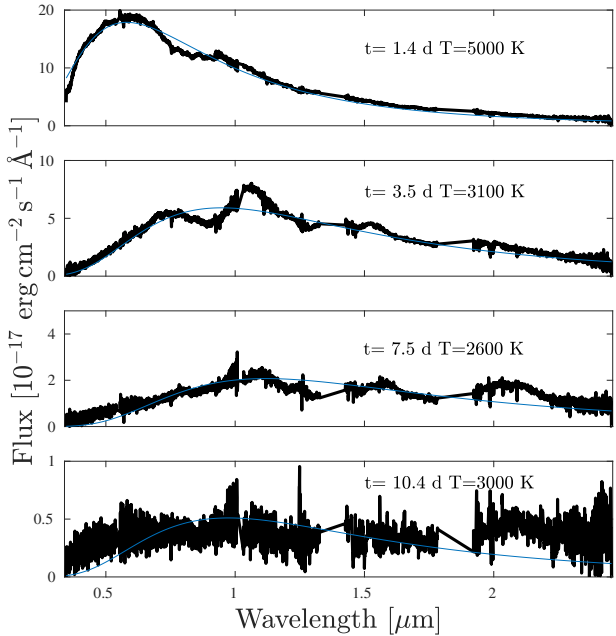}}
\caption{Examples for blackbody fits to the spectroscopic data.
Solid lines represent the best fit blackbody fit.
Time after the merger and best fit temperature are indicated on each plot. See also Table~\ref{tab:BolLumSpec}.
\label{fig:SpecSpectBB}}
\end{figure}
\begin{figure*}
\centerline{\includegraphics[width=16cm]{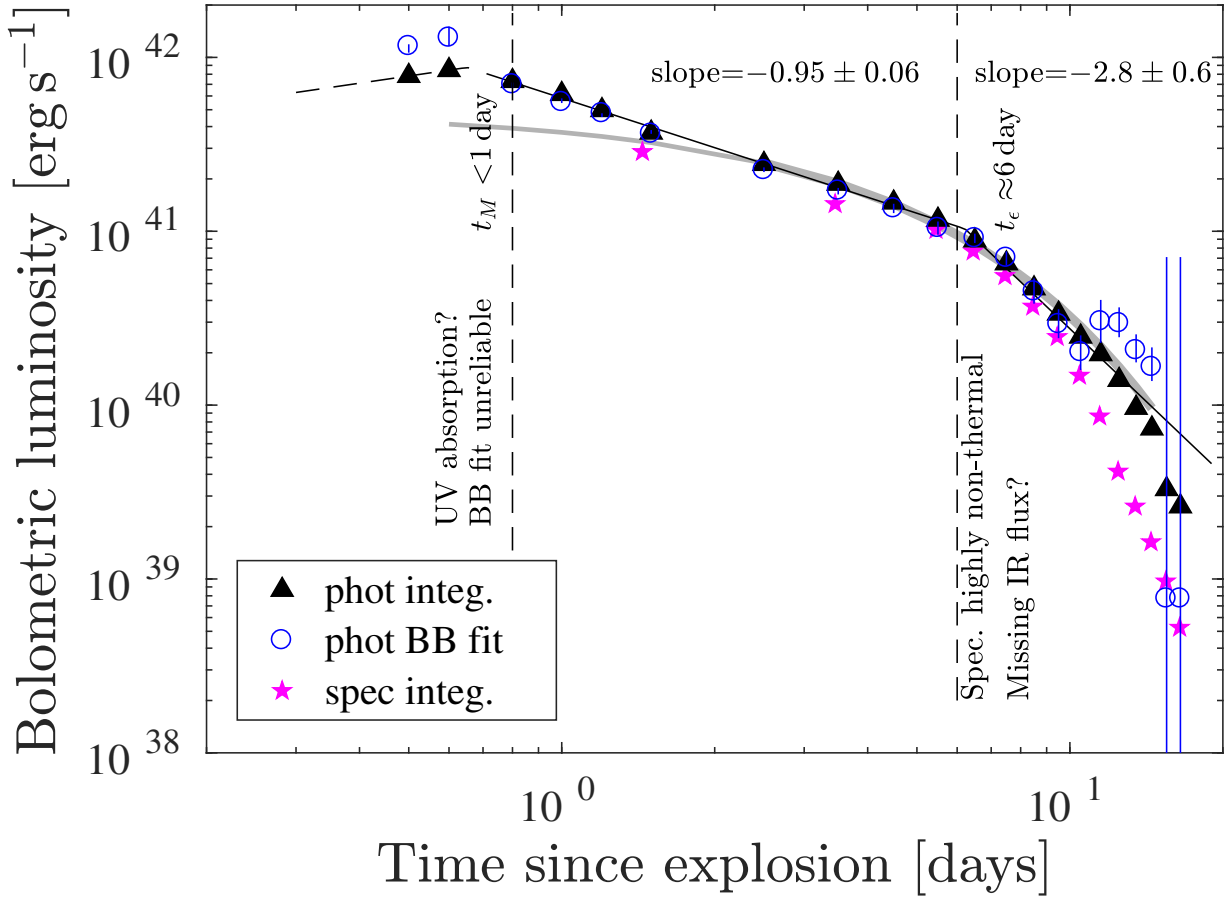}}
\caption{The estimated bolometric light curve of GW170817/AT\,2017gfo.
The blue-empty circles show the estimates based on fitting blackbody spectra to the photometric points.
The filled triangles are based on
trapezoidal integration of the photometric data points, while the filled stars represent the integration of the X-Shooter spectra.
The bolometric light curve is highly uncertain after $t\approx7$\,day, as
at this stage we cannot rule out that we miss some of the emission
in the IR.
The solid line shows a broken power-law fit
to the trapezoidal integration of the photometric data points-based light curve,
with parameters given in Table~\ref{tab:pow_fit}.
The thick-solid gray line shows the best-fit exponential to the data between $t=2$ to $t=15$\,days,
while the thin-solid gray lines shows the extension of this fit before $t=2$\,days.
The best-fit exponential has a decay timescale of $3.7\pm0.3$\,days.
\label{fig:LumBolFit}}
\end{figure*}
\begin{figure}
\centerline{\includegraphics[width=8cm]{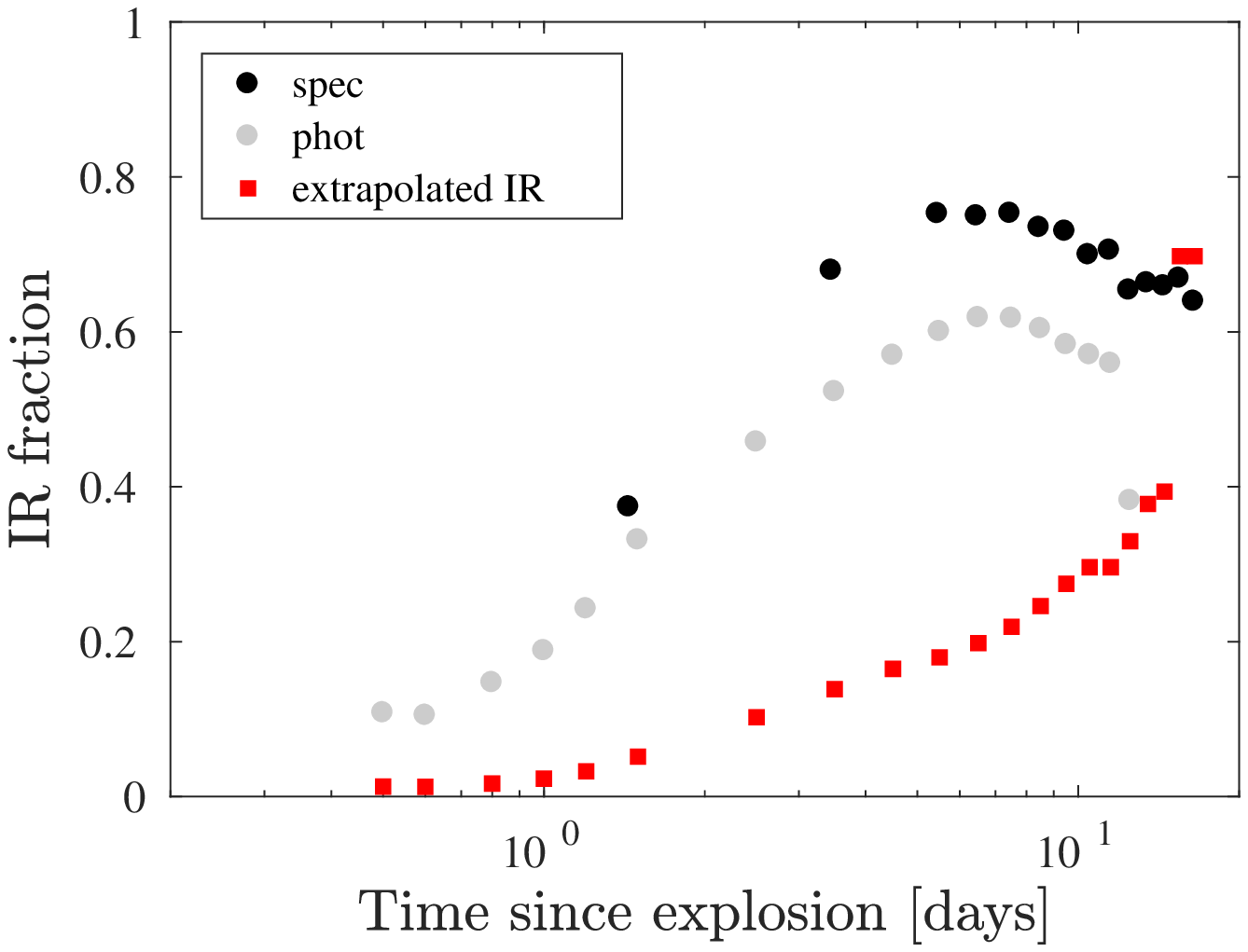}}
\caption{The fraction of the luminosity emitted in the IR
($1-2.45$$\mu$m) compared with the entire observed luminosity.
Black points represent the estimates based on the integration
of the spectroscopic data, and gray points represent the trapezoidal
integration of the photometric data without the Rayleigh-Jeans extrapolated tail. The red squares show the fraction of the extrapolated Rayleigh-Jeans tail.
We note that the spectra taken after $t=10$\,days are very noisy
and may suffer from background subtraction and calibration issues.
\label{fig:IRfrac}}
\end{figure}

We estimated the bolometric luminosity from the photometric data using
two methods.
First, we corrected the apparent magnitude of each band to
the Galactic extinction in the direction of the transient
($E_{\rm B-V}=0.11$\,mag; \cite{1998ApJ...500..525S};
\cite{1989ApJ...345..245C}), converted it to flux,
and integrated the flux using trapezoidal integration.
We assumed that above the highest wavelength
measurement, the spectrum falls like a Rayleigh-Jeans tail ($\lambda^{-4}$),
and added this part to the integration.
This Rayleigh-Jeans tail contributes
to the estimated bolometric luminosity
between
$\simeq1$\% at 1\,day after the merger to about 18\% at 5\,days,
25\% at 9\,days, and $\sim70$\% at an age of 16\,days.

Second, we fitted a blackbody (with $E_{\rm B-V}=0.11$\,mag redenning)
to the photometric points and their errors.
The $\chi^{2}$ of the fits were typically much larger than the number of degrees of freedom, indicating on the large deviations from blackbody
spectrum.
Therefore, we renormalized the photometric errors such that the
$\chi^{2}$ will be equal to the number of degrees of freedom.
This step does~not change much the best fit values but it
is crucial for a proper estimation of the uncertainties.
This fit provides the effective temperature, and photospheric radius.
Since the errors in the photospheric radius and temperature
are highly correlated, we estimated the errors in temperature,
radius and luminosity by marginalizing the likelihood over all other parameters.
We present a sample of the blackbody fits to the photometric data
in Figure~\ref{fig:PhotSpectBB}.
Table~\ref{tab:BolLum} in \S~\ref{sec:data} presents our bolometric luminosity based on
the trapezoidal integration, and blackbody fit
as well as the best fit temperature and radius.

We estimated the bolometric luminosity also by
integrating the spectra.
For that we used only the X-Shooter spectra
(\cite{2017Natur.551...67P}; \cite{2017Natur.551...75S}).
Before the integration, we re-calibrated the spectra by setting its synthetic photometry in some band to match the photometric observations.
For $t<5$\,days we used the synthetic photometry in the $r$-band,
while for $t>5$\,days we used the {\it HST} $F110W$ band.
We corrected all the spectra for Galactic extinction.
We also fitted a blackbody to the spectra, and several of the fits
are shown in Figure~\ref{fig:SpecSpectBB}.
The best-fit blackbody to the spectra are summarized in Table~\ref{tab:BolLumSpec} in \S~\ref{sec:data}.

Figure~\ref{fig:LumBolFit} presents the three estimates of the bolometric luminosity as a function of time. The various estimates are consistent between day one and day ten. However, they are inconsistent at early times ($t\ltorder1$\,day), and late times ($t\gtorder10$\,day).  At late times, figures~\ref{fig:PhotSpectBB} and \ref{fig:SpecSpectBB} suggest that the spectrum is not well described by a blackbody. At early times ($t\ltorder1$\,day), the blackbody fit is reasonable. However, the best blackbody fit underestimates the UV data points. One possible explanation is that the UV emission is affected by UV absorption features, and at these early times, the measured effective temperature is underestimated. Given these considerations we adopted, as our best estimate bolometric luminosity, the trapezoidal integration of the photometric data. The radiated energy (integrated luminosity) from $t\approx0.5$\,days to $t=16.5$\,days is about $1.5\times10^{47}$\,erg.
As seen in figures~\ref{fig:PhotSpectBB} and~\ref{fig:SpecSpectBB}, at early times ($t\ltorder5$\,days) the spectrum is
well represented by a blackbody. Hence, it is optically thick and the effective temperature estimates are reliable. However, at late times ($t\gtorder10$\,days), the spectrum shows considerable deviations from a blackbody spectrum. {A reasonable and consistent  explanation is that at $t \gtorder 7$\,days the emission is optically thin (an alternative explanation, of a black body with very broad absorption lines, cannot be excluded but is challenged by the presence of emission lines at the blue region of the spectrum).} Therefore, at these late times the blackbody fits cannot be trusted for an estimation of the effective temperature and bolometric luminosity. This also suggests that some of the spectral features in the late time spectra are in fact emission lines.

An important clue to the nature of the ejecta in AT\,2017gfo may come from the broad-band photometry of the event. The reason is that the opacity of some elements may be very sensitive to wavelength (see discussion in \S\ref{sec:late}). Figure~\ref{fig:IRfrac} shows the fraction of luminosity emitted in the IR ($1-2.45$$\mu$m) compared with the entire observed luminosity. Black points represents the estimates based on the integration of the spectroscopic data, and gray points represents the trapezoidal integration of the photometric data without the Rayleigh-Jeans extrapolated tail. The red squares shows the fraction of the extrapolated Rayleigh-Jeans tail.

\subsection{Simple fits to the bolometric light curve, effective temperature and photospheric radius}
\label{sec:fits}

Next, we fitted a broken power-law to the bolometric light curve
based on the trapezoidal integration of the photometry.
The best fit broken power law is presented in Figure~\ref{fig:LumBolFit}.
We find that the data are described well by a power-law with two
breaks. The best fit values are given in Table~\ref{tab:pow_fit}
{\citep[the inferred indices are consistent with those obtained in the analysis of][]{Arcavi18}}.
We also attempted to fit an exponential decay to the light curve,
and the best fit is shown as the dashed line in the figure.
The implications are discussed in detail in \S~\ref{sec:comp_obs}.
\begin{figure}
\centerline{\includegraphics[width=8cm]{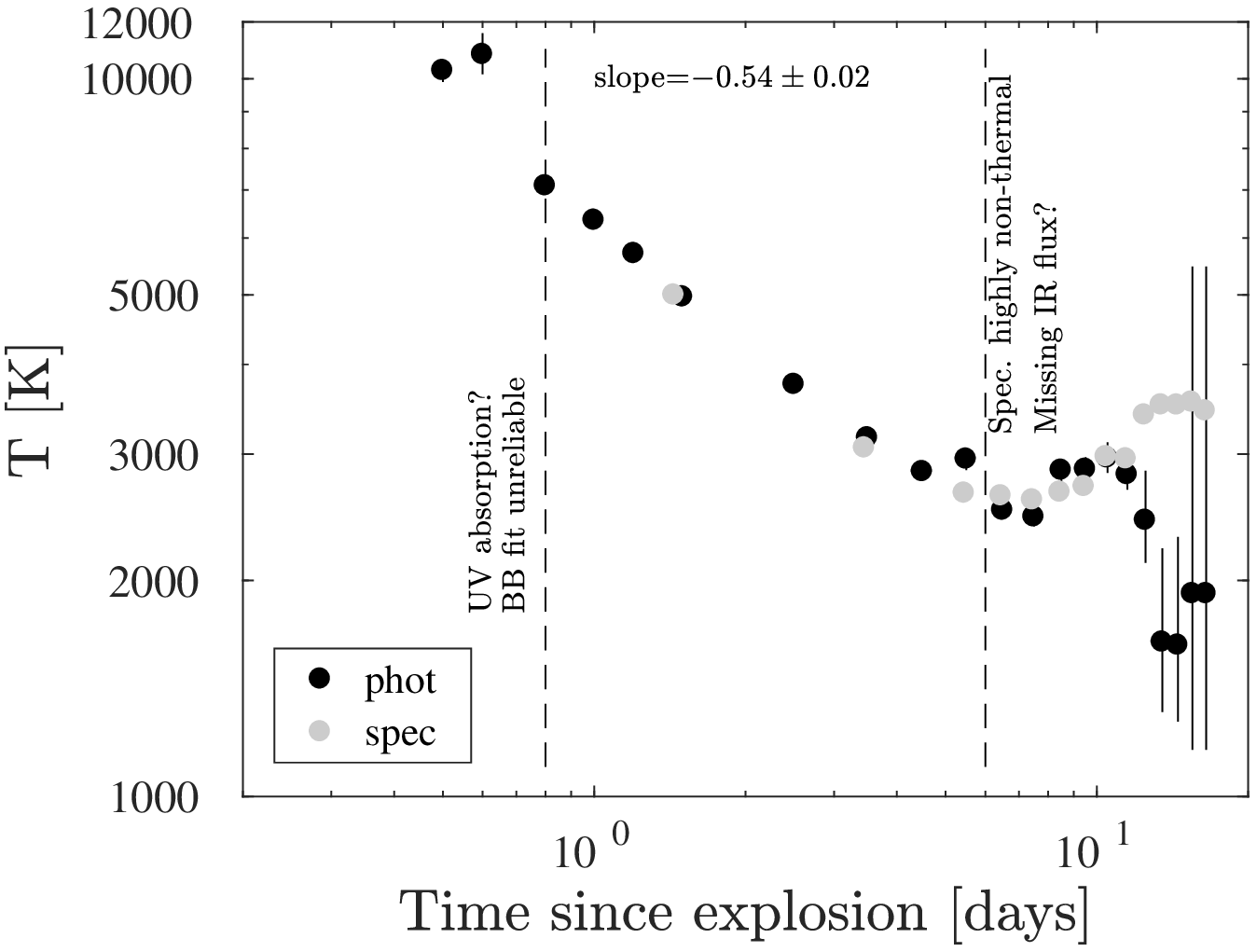}}
\caption{The effective temperature as estimated from fitting
the multi-band photometric data (black circles) and spectroscopic data
(gray circles) with a blackbody spectrum.
The estimates are unreliable at early times ($\ltorder1$\,day) and at late times ($\gtorder7$\,day),
when the UV photometric points deviate from a blackbody fit.
The best-fit power-law index, between day 1 to 5, is $-0.54\pm0.02$.
\label{fig:T}}
\end{figure}
\begin{figure}
\centerline{\includegraphics[width=8cm]{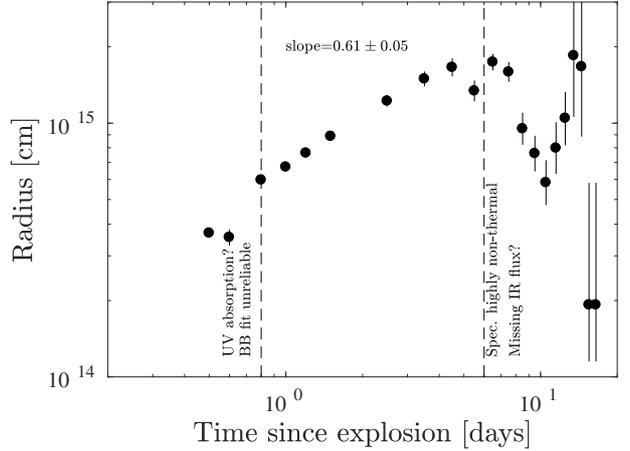}}
\caption{The same as Figure~\ref{fig:T}, but for the photospheric radius.
The best-fit power-law index, between day 1 to 5, is $0.61\pm0.05$.
\label{fig:R}}
\end{figure}
\begin{figure}
\centerline{\includegraphics[width=8cm]{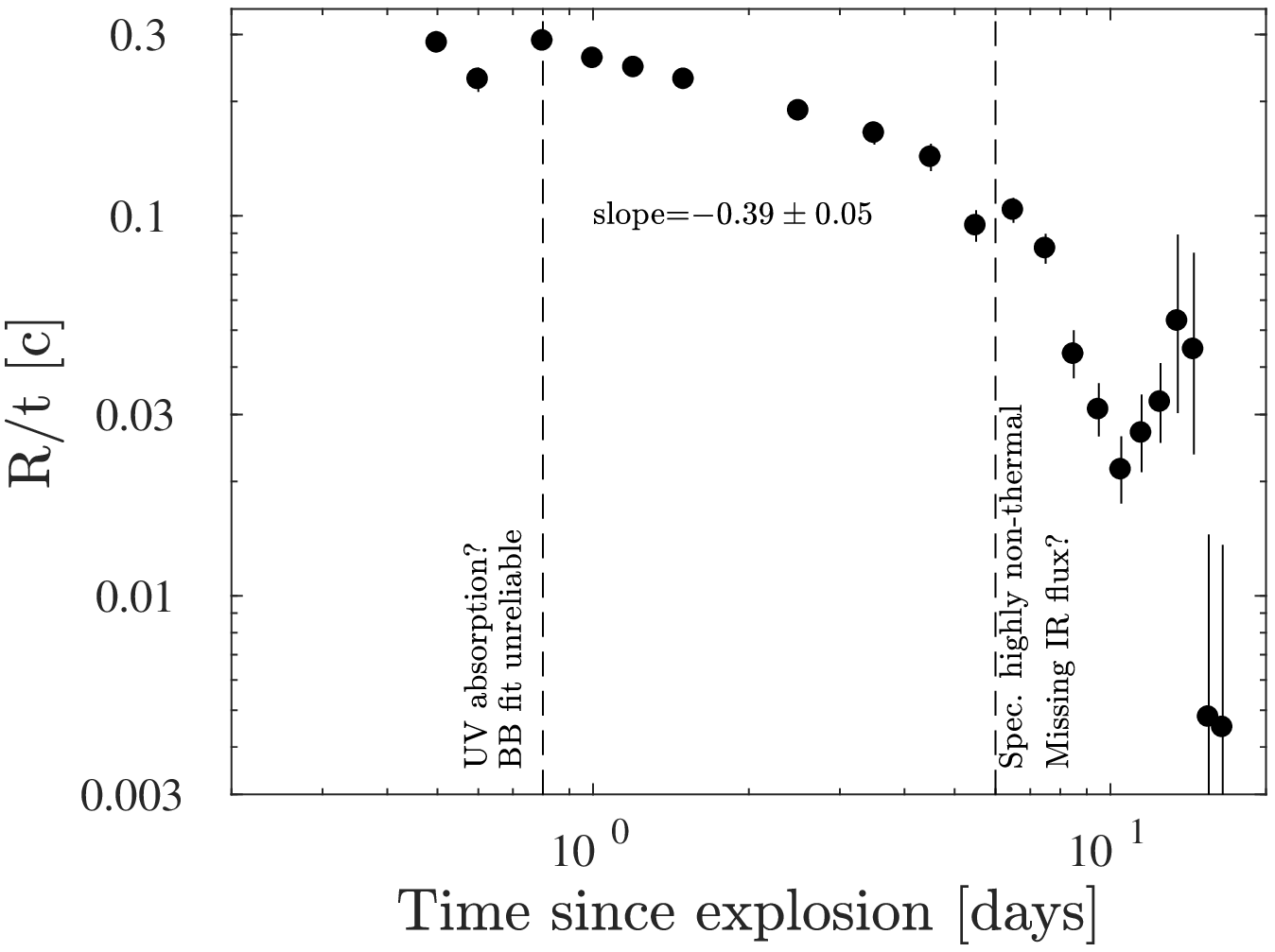}}
\caption{The same as Figure~\ref{fig:T}, but for the photospheric radius divided by time since merger.
The best-fit power-law index, between day 1 to 5, is $-0.39\pm0.05$.
\label{fig:V}}
\end{figure}

{All power-law fits were done by minimizing the $\chi^{2}$ and using the individual errors of the data points. In all the cases we found that the probability to get the $\chi^2/$dof is larger than about 5~\%. Therefore, we did~not rescale the measurement errors. Furthermore, the reported errors are the usual $1~\sigma$ errors without marginalization over all other parameters.
}
In the broken-power law description,
the most robust features of the light curve are as follows.
(i) At $t=1$\,day the bolometric luminosity is about $6\times10^{41}$\,erg\,s$^{-1}$.
(ii) Between $t>1$\,day and $t=6.2\pm0.7$\,day the
bolometric light curve is well represented by a power-law decay
with a power-law index of $-0.95\pm0.06$ {($\chi^{2}/$dof$=$17.2/15 for the entire broken power-law fit)}.
(iii) At $t=6.2\pm0.7$\,day there is some indication for a break in
the light curve, followed by a steeper decay rate.
We cannot rule out however the possibility that at these times the emission shifted into the mid IR (i.e., above 2.4$\mu$m).
(iv) If the break at $t=6.2\pm0.7$\,day is not a consequence
of missing IR flux, then the best power-law index
after this break is $-2.8\pm0.6$.
We note that the existence of this break can be tested using
mid-IR observations (e.g., using Spitzer).
(v) {There is an indication of a break in the bolometric light curve at $t\ltorder 1$\,day.
However, given the uncertainty of the bolometric luminosity at these early times,
we do not think that the existence of the break is robust. Therefore, at this stage
all we can claim is that there if there is break, it is at t<1 day.}
\begin{deluxetable}{ll}
\tablecolumns{4}
\tablewidth{0pt}
\tablecaption{Bolometric luminosity power-law fit}
\tablehead{
\colhead{Parameter}          &
\colhead{Value}
}
\startdata
  First break & $t_1<1$~d \\
  Second break & $t_2=6.2\pm0.7$~d \\
  $t_1<t<t_2$ power-law index & $-0.95\pm0.06$ \\
  $t>t_2$ power-law index & $-2.8\pm0.6$ \\
  $L(1{\rm d})$ & $6.1_{-0.1}^{+0.3}\times10^{41}{\rm erg/s}$
\enddata
\label{tab:pow_fit}
\end{deluxetable}

Figure~\ref{fig:T} presents the blackbody temperature estimated from the photometric data and spectroscopic data, while Figure~\ref{fig:R} shows the photospheric radius as a function of time
as derived from the blackbody fits.

Again, these fits are reliable only between day one and about day five.
After about one week it seems that the spectral energy distribution
of AT\,2017gfo is poorly described by a blackbody radiation,
and therefore the estimated effective temperature and photospheric
radius are unreliable.
We find that between day one to five, the temperature evolution
is well described by a power-law with an index of $-0.54\pm0.02$ {($\chi^{2}/$dof$=$2.3/3)},
while the radius is well described by a power-law with an index
of $0.61\pm0.05$ {($\chi^{2}/$dof$=$0.4/3)}.

Figure~\ref{fig:V} shows the estimated
photospheric velocity calculated by $R/t$.
Between day one to day five, the velocity is consistent with
a power-law decay, between day 1 to 5, with index of $=-0.39\pm0.05$ {($\chi^{2}/$dof$=$0.4/3)}.
This plot suggests that the velocity at day one is of the order
of $0.3c$.

\section{A simple model for the emission of radiation from the expanding ejecta}
\label{sec:Model}

The model and its underlying assumptions are described in \S~\ref{sec:ModelDesc}. The main model results are given in \S~\ref{sec:ModelSum}. These include the multi-band and bolometric light curves, $L(t)$, photospheric radii, $r_{\rm ph}(t)$, and effective temperatures, $T_{\rm eff}(t)$, predicted by the model. The derivation of these results is given in the appendix, \S~\ref{sec:ModelDer}.

\subsection{Model description}
\label{sec:ModelDesc}

The smooth continuous evolution of the luminosity, temperature and photospheric radius at $t\lesssim6$~d motivates us to consider a simple model, in which the {microscopic} properties of the emitting plasma {(e.g. composition, opacity)} do not vary abruptly across the ejecta. The fact that the photospheric radius growth is slower than $r\propto t$ suggests that the photosphere gradually penetrates into slower parts of the ejecta, and the power-law evolution, $r_{\rm ph}\propto t^{0.6}$, suggests a power-law distribution of velocity as a function of mass. We therefore consider a spherical ejecta with a velocity profile
\begin{equation}
	v(m) = v_{\rm M}(\tilde{m}/M)^{-\alpha} \equiv v_{\rm M} m^{-\alpha},
\label{eq:vm}
\end{equation}
where $\tilde{m}$ is the integrated mass measured from the outer edge (i.e., highest velocity), $\tilde{m} = \int_{r}^{\infty}{4\pi r^{2}\rho(r)dr}$, $M$ is the total ejecta mass, $m$ is the fractional mass, $m=\tilde{m}/M$, and $v_{\rm M}$ is the characteristic velocity of the slowest parts of the ejecta. Assuming that Eq.~(\ref{eq:vm}) holds over (at least) a factor of a few in $m$, and given that $v_{\rm M}$ is not far below $c$, physically meaningful values of $\alpha$ are limited to $\alpha\lesssim2$ (as otherwise the fastest velocity would exceed $c$).

We assume a uniform opacity, $\kappa$, which is allowed to evolve with time,
\begin{equation}
	\kappa = \kappa_{\rm M}(t/t_{\rm M})^\gamma.
\label{eq:kappa_t}
\end{equation}
Here $\kappa$ is the effective opacity governing the photon diffusion time, taking into account the velocity gradient ("expansion opacity") and averaged over frequency. The temporal evolution of the opacity is allowed in order to account for variations due to the temperature and density evolution of the plasma. Moreover, it may partially capture the effect of an opacity gradient due to a composition gradient within the ejecta, since the opacity at the region from which photons are emitted would evolve with time in this case. We note, however, that this simple parametrization cannot capture large abrupt variations of the opacity, due to strong spatial or temporal evolution. {In particular, a rapid increase of the opacity due to a transition between regions of different compositions (e.g. low to high Lanthanide fraction) cannot be accounted for. Similarly,} a rapid drop of opacity due to recombination, which is expected at a temperature of $\sim3000$~K for both Iron group and Lanthanide elements \citep[e.g.][]{Kasen13}, cannot be accounted for. The implications of this rapid drop are discussed in \S~\ref{sec:ModelSum}.

We define $t_{\rm M}$ as the time at which photons begin to escape, on a dynamical time scale, from the slowest part of the ejecta. At any time $t$, photons escape from regions with optical depth smaller than $c/v$ (where the photon diffusion time is shorter than $t$). For the power-law velocity profile we adopted, the optical depth beyond radius $r$ is
\begin{eqnarray}\label{eq:tau}
    \tau &=& \int_r^\infty dr\kappa\rho=\kappa\int_r^\infty dr \frac{-1}{4\pi r^2}\frac{d\tilde{m}}{dr}
    \nonumber \\
    &=&\frac{\kappa M}{4\pi t^2}\int_0^m dm v^{-2}=\frac{\kappa M}{4\pi (1+2\alpha)v_{\rm M}^2t^2}m^{2\alpha+1}
\end{eqnarray}
(note that $\tau\propto r^{-(1+2\alpha)/\alpha}$ and $\rho\propto r^{-(1+3\alpha)/\alpha}$), and
\begin{equation}\label{eq:tM}
  t_{\rm M}^2\equiv \frac{1}{4\pi(1+2\alpha)}\frac{\kappa_{\rm M} M}{c v_{\rm M}}.
\end{equation}
We refer to the location within the ejecta at which $\tau=c/v$ as the "diffusion sphere". At $t=t_{\rm M}$ the diffusion sphere reaches $m=1$, and at $t>t_{\rm M}$ photons escape from the entire ejecta on a time scale shorter than $t$.

We assume a uniform radioactive energy release per unit mass,
\begin{equation}\label{eq:Edot}
   \dot{\varepsilon}=\dot{\varepsilon}_{\rm M} (t/t_{\rm M})^{-\beta}.
\end{equation}
This temporal behavior represents the decay of nuclei with a wide range of decay times, as expected to be produced in the expansion of the neutron rich ejecta \citep{LattimerSchramm74,LiPac98}. Assuming a uniform distribution in logarithmic life time intervals, $\beta\sim1$ was suggested by \citet{LiPac98}. Detailed numerical calculations find $1.1\lesssim\beta\lesssim1.4$ and $\dot{\varepsilon}(1{\rm d})\approx 2\times10^{10}{\rm erg\,g^{-1}s^{-1}}$ (to within a factor of a few) for $Y_{\rm e}<0.4$ \citep[e.g.][]{Freiburghaus99,Metzger10,Goriely11,Roberts11,Korobkin12,Wanajo14,Lippuner15,Barneskasen16,Hotokezaka16heat,Rosswog17}. A uniform radioactive energy release is likely to be a good approximation, since the energy release obtained for a wide range of $Y_{\rm e}<0.4$ is nearly independent of $Y_{\rm e}$.

The energy deposited in the plasma is a fraction of the total radioactive energy release. The energy carried by neutrinos is not deposited. $\gamma$-rays also escape at relatively early time, as demonstrated by the following argument. At the place from which optical/UV photons escape, $\kappa\rho r\sim c/v$, the optical depth for $\gamma$-rays is $\kappa_\gamma\rho r\approx (c/v)\kappa_\gamma/\kappa$, where the $\gamma$-ray opacity is $\kappa_\gamma\approx10^{-1.5}{\rm cm^2/g}$ for $\gamma$-rays in the relevant energy range, $\gtrsim 1$~MeV {\citep{Longair92Book}}. Since a few scatterings are required for energy deposition (i.e. for significant energy loss of the $\gamma$-ray photon in Compton scatterings), the fraction of the $\gamma$-ray energy deposited is $\approx(c/3v)\kappa_\gamma/\kappa\approx 10^{-1.5}(\kappa/1{\rm cm^2g^{-1}})^{-1}\ll1$. For our simple model, we therefore consider only the energy deposited by electrons and positrons, assuming that most of the energy is generated by $\beta$-decays,
\begin{equation}\label{eq:Edepo}
  \dot{\varepsilon}_d(t,m)=(1-f_{\nu\gamma})f_{ed}(m)\dot{\varepsilon},
\end{equation}
where $f_{\nu\gamma}$ is the energy fraction carried by neutrinos and $\gamma$-rays, and $f_{ed}(m)$ is the $m$-dependent fraction of e$^\pm$ energy deposited in the plasma. {We note that the assumption that the energy generation is dominated by beta-decay is generally valid, since the energy deposited by fission products is sub-dominant to that deposited by photons and electrons, and the energy deposited by alpha-particles is less than a few percent of the electron-positron deposition up to 10~d \citep[e.g.][]{Barneskasen16}.}

In the relevant energy range, $\sim0.5-2$~MeV, the e$^\pm$ energy loss is well described by $dE/dX\approx 1{\rm MeV}/{\rm g\,cm^{-2}}$ \citep[within a factor of a few for a wide range of materials, e.g.][]{Longair92Book}. Assuming that the e$^\pm$ are confined to the plasma (which would be valid even for very low magnetic fields), the fraction of the energy they lose over time $t$ is
\begin{equation}\label{eq:fed}
  f_{ed}\approx \min\left[1,\rho c t E^{-1} dE/dX\right] = \min[1,\kappa_e\rho ct]
\end{equation}
with an effective electron "opacity", $\kappa_e\approx1{\rm cm^2/g}$ (if electrons escape freely, over time $r/c$, the deposited fraction would be smaller by $v/c$,  $f_{ed}\approx \kappa_e\rho r$, which is a small correction for our $v/c\sim1$). At late time, when $f_{ed}<1$ at all parts of the ejecta, the energy deposition in the entire ejecta decreases like $1/t^2$, since $f_{ed}\propto \rho t\propto t^{-2}$. The mass averaged value of $f_{ed}$, $\int dm f_{ed}=\kappa_e ct\int dm\rho$ (recall $\rho\propto m^{1+3\alpha)}$), is given at late time by $(t/t_\varepsilon)^{-2}$, with
\begin{equation}\label{eq:tE}
  t_\varepsilon^2=\frac{1}{\alpha(2+3\alpha)}\frac{c\kappa_e M}{4\pi v_{\rm M}^3},\quad \kappa_e\approx1{\rm cm^2/g}.
\end{equation}
The fact that $\kappa_e\approx1{\rm cm^2/g}$ implies that for low opacity, $\kappa_{\rm M}\lesssim1{\rm cm^2/g}$, the time at which the ejecta becomes optically thin, $t_{\rm ph}$, is similar to $t_\varepsilon$. For $\kappa_{\rm M}\gg1{\rm cm^2/g}$, $t_\varepsilon\ll t_{\rm ph}$.

The following point is important to clarify. If electrons and positrons are confined to the plasma, they lose energy adiabatically (i.e to the kinetic energy of the accelerated plasma) as the plasma expands, $dE/dt\approx -x E/t$ with $x=1(2)$ for highly (non)-relativistic particles. Since the particles lose most of their energy adiabatically over a time $t$, the fraction of their energy that is converted to thermal energy is the fraction that they lose by ionization/Bremsstrahlung interactions over time $t$, $\rho c t E^{-1} dE/dX$. Neglecting adiabatic losses, the confined electrons never lose much of their energy and they accumulate in the plasma over time: without adiabatic losses we have $dE/dt\propto\rho\propto t^{-3}$, which implies that e$^\pm$ produced at $t>t_\varepsilon$ lose only a small fraction of their energy $\approx (t/t_\varepsilon)^{-2}$ (at all times). For a radioactive energy release rate $\propto t^{-1}$, neglecting adiabatic losses leads to a logarithmic divergence of the energy carried by confined high energy e$^\pm$, $\epsilon\propto\ln(t/t_\varepsilon)$, which in turn leads to a non-physical over-estimate of the rate of energy deposition at late times \cite[this is the case, e.g., in the analytic estimates of][see their Eqs.~29, 31 \&~32]{Barneskasen16}.

\subsection{Summary of the main model results}
\label{sec:ModelSum}

The numerical values of $t_{\rm M}$ and $t_\varepsilon$ are give by
\begin{eqnarray}\label{eq:tM_num}
  t_{\rm M}&\equiv& \left[\frac{1}{4\pi(1+2\alpha)}\frac{\kappa_{\rm M} M}{c v_{\rm M}}\right]^{1/2}
  \nonumber \\
  &=& 1.5\frac{1}{(1+2\alpha)^{1/2}}\left(\frac{\kappa_{\rm M}}{1\rm cm^2/g} \frac{M}{0.01M_\odot}\frac{0.1c}{v_{\rm M}}\right)^{1/2}\,{\rm d},
\end{eqnarray}
\begin{eqnarray}\label{eq:t_E_num}
  t_\varepsilon&\equiv&\left[\frac{c\kappa_e M}{4\pi\alpha(2+3\alpha) v_{\rm M}^3}\right]^{1/2}
  = 15\frac{1}{\alpha^{1/2}(2+3\alpha)^{1/2}} \nonumber \\
  &\times&\left(\frac{\kappa_e}{1\rm cm^2/g} \frac{M}{0.01M_\odot}\right)^{1/2}\left(\frac{0.1c}{v_{\rm M}}\right)^{3/2}\,{\rm d}.
\end{eqnarray}
The ratio of these two times is given by
\begin{equation}\label{eq:XM}
  X_{\rm M}\equiv\left(\frac{t_{\rm M}}{t_\varepsilon}\right)^2=
  \frac{\alpha(2+3\alpha)}{1+2\alpha}\left(\frac{v_{\rm M}}{c}\right)^2\frac{\kappa_{\rm M}}{\kappa_e}.
\end{equation}

The qualitative behavior of the bolometric light curves depends on the relation between $t_{\rm M}$ and $t_\varepsilon$. For $t_{\rm M}<t_\varepsilon$ ($X_{\rm M}<1$)
\begin{equation}
\label{eq:L_tM_tE2}
  L=L_{\rm M}
  \left\{
    \begin{array}{ll}
      \eta X_{\rm M}^{-1+\beta/2}(t/t_{\rm M})^{\frac{2-\gamma}{1+\alpha}-\beta+s(1-\frac{\beta}{2})}, &
             \hbox{$t<t_{e D}=X_{\rm M}^{1/s}t_{\rm M}$ and $s>0$;} \\
      \eta(t/t_{\rm M})^{\frac{2-\gamma}{1+\alpha}-\beta}, &
             \hbox{$t<t_{\rm M}$ (and $t_{e D}<t$ for $s>0$);} \\
      (t/t_{\rm M})^{-\beta}, & \hbox{$t_{\rm M}<t<t_\varepsilon=X_{\rm M}^{-1/2}t_{\rm M}$;} \\
      X_{\rm M}^{\beta/2}(t/t_\varepsilon)^{-\beta-2}, & \hbox{$t_\varepsilon<t$,}
    \end{array}
  \right.
\end{equation}
and for $t_{\rm M}>t_\varepsilon$ ($X_{\rm M}>1$)
\begin{equation}
\label{eq:L_tE_tM2}
  L=L_{\rm M}
  \left\{
    \begin{array}{ll}
      \eta(t/t_{\rm M})^{\frac{2-\gamma}{1+\alpha}-\beta}, &
             \hbox{$t<t_{e D}=X_{\rm M}^{1/s}t_{\rm M}$ and $s<0$;} \\
      \eta X_{\rm M}^{-1+\beta/2}(t/t_{\rm M})^{\frac{2-\gamma}{1+\alpha}-\beta+s(1-\frac{\beta}{2})}, &
        \hbox{$t<t_{\rm M}$ (and $t_{e D}<t$ for $s<0$);} \\
      X_{\rm M}^{-1}(t/t_{\rm M})^{-\beta-2}, & \hbox{$t_{\rm M}<t$.}
    \end{array}
  \right.
\end{equation}
Here,
\begin{eqnarray}\label{eq:LM_num}
  L_{\rm M}&=&M(1-f_{\nu\gamma})\dot{\varepsilon}_{\rm M}=2\times10^{41}
  \nonumber \\
  &\times&\frac{M}{0.01M_\odot}\frac{(1-f_{\nu\gamma})\dot{\varepsilon}(t=1{\rm d})}{10^{10}{\rm erg/g\,s}}
  \left(\frac{t_{\rm M}}{1\rm d}\right)^{-\beta}{\rm erg/s},
\end{eqnarray}
\begin{equation}\label{eq:s_eta}
  s=\frac{4\alpha-(1+3\alpha)\gamma}{(1+\alpha)},\quad
  \eta= 1+\frac{2-\gamma}{(1+\alpha)(2-\beta)}.
\end{equation}
$t_{eD}$ is the time at which the location of the diffusion sphere, where $\tau=c/v$, coincides with the location in the ejecta beyond which the fraction of e$^\pm$ energy deposited in the ejecta drops below unity.

Eqs.~(\ref{eq:L_tM_tE2}) and~(\ref{eq:L_tE_tM2}) were derived assuming that the velocity distribution, $v\propto m^{-\alpha}$, corresponding to $m\propto v^{-1/\alpha}$, extends to arbitrarily large velocity (small mass). In reality, we expect $m(v)$ to be strongly suppressed above some characteristic (maximal) velocity $v_{\rm max}$. Eqs.~(\ref{eq:L_tM_tE2}) and~(\ref{eq:L_tE_tM2}) hold therefore at times when the diffusion sphere lies at masses $m$ for which $v(m)<v_{\rm max}$, i.e at times greater than
\begin{equation}\label{eq:t_b}
  t_b=\left(\frac{v_{\rm M}}{v_{\rm max}}\right)^{\frac{(1+\alpha)}{(2-\gamma)\alpha}}t_{\rm M}
\end{equation}
(see Eq.~\ref{eq:m_diff}). At earlier time, the emission is produced by the outer part of the fastest shell of the ejecta, with optical depth $c/v_{\rm max}$. For the case where the density in this region is large enough at $t<t_b$ to allow efficient e$^\pm$ energy deposition (which is the case for $X_{\rm M}\ll1$), the luminosity at $t<t_b$ follows $L\propto t^{2-\gamma-\beta}$.

The time $t_{\rm ph}$ at which the photosphere crosses the entire ejecta (i.e., after which entire ejecta is optically thin) is
\begin{equation}
t_{\rm ph} = t_{\rm M} \left(\frac{c}{v_{\rm M}}\right)^{1/(2-\gamma)}.
\label{eq:t_ph}
\end{equation}
For $t<t_{\rm ph}$ the photospheric radius is given by
\begin{equation}\label{eq:r_phot}
  \frac{r_{\rm ph}(t)}{ct_{\rm M}}=
  \left(\frac{v_{\rm M}}{c}\right)^{\frac{1+\alpha}{1+2\alpha}}
  \left(\frac{t}{t_{\rm M}}\right)^{\frac{1+\gamma\alpha}{1+2\alpha}}.
\end{equation}
Here too, Eq.(\ref{eq:r_phot}) holds at times at which the photosphere lies at masses $m$ for which $v(m)<v_{\rm max}$. At earlier time, $t<(c/v_{\rm max})^{1/(2-\gamma)} t_b$ (see Eq.~\ref{eq:m_phot}), the photosphere lies at the outer part of the fastest moving ejecta, i.e. $r_{\rm ph}=v_{\rm max}t$.

The effective temperature may be derived from $L$ and $r_{\rm ph}$.
\begin{eqnarray}\label{eq:TM_num}
  T_{\rm M}&\equiv& T_{\rm eff.}(t=t_{\rm M})
  =\left[\frac{L_{\rm M}}{4\pi\sigma r^2_{\rm ph}(t_{\rm M})}\right]^{1/4} \nonumber \\
  &=&0.47(1+2\alpha)^{3/8}\left(\frac{v_{\rm M}}{c}\right)^{\frac{\alpha}{2(1+2\alpha)}}\left(\frac{t_{\rm M}}{1~\rm d}\right)^{\frac{1-\beta}{4}}\left(\frac{\kappa_{\rm M}}{1\rm cm^2/g}\right)^{-3/8}  \nonumber \\
  &\times&
  \left[\frac{(1-f_{\nu\gamma})\dot{\varepsilon}(t=1{\rm d})}{10^{10}{\rm erg/g\,s}}\right]^{1/4}
  \left(\frac{\rm M}{0.01M_\odot}\frac{v_{\rm M}}{0.2c}\right)^{-1/8}
  {\rm eV}.
\end{eqnarray}

At $t>t_{\rm ph}$ the ejecta is optically thin, and the emission is roughly given by
\begin{equation}\label{eq:Lnu_thin}
  L_\nu\approx\min\left[1,\pi R^2/(\kappa_\nu M) \right]\kappa_\nu M c u^{BB}_\nu(T),
\end{equation}
where $\kappa_\nu$ is the opacity and $u^{BB}_\nu(T)$ is the blackbody spectral energy density at the plasma temperature $T$, which is determined by
\begin{equation}\label{eq:Tthin}
  \dot{\varepsilon}_d=\int d\nu L_\nu(T).
\end{equation}
The opacity that we have used so far, $\kappa_{\rm M}$, is the average (expansion corrected) opacity, while $\kappa_\nu$ is the frequency dependent (expansion corrected) opacity. The following point is important to note. At $t<t_{\rm ph}$ the radiation energy density is close to the blackbody energy density, and the plasma may be assumed to be close to local thermal equilibrium (LTE). At $t>t_{\rm ph}$ the radiation energy density drops below the blackbody energy density (corresponding to the plasma temperature $T$), and the opacities may differ significantly from their LTE values.

As mentioned in the preceding section, the rapid drop of opacity due to recombination, which is expected at a temperature of $\sim3000$~K for both Iron group elements and Lanthanides \citep[e.g.][]{Kasen13}, is not accounted for in the above derivation. This rapid drop in opacity implies that the temperature cannot decrease significantly below the recombination temperature. In cases where this temperature is reached before the ejecta is transparent, i.e. at $t<t_{\rm ph}$, the photosphere will recede to the point where the plasma is ionized, while if this temperature is reached after transparency, the temperature will be kept close to recombination in order to enable the generation of radiation that is required in order to release the deposited energy.

\section{Applying the model to GW170817}
\label{sec:comp_obs}

In this section we show that the UV to IR emission of GW170817 is consistent with the simple model described in the preceding section (see figures~\ref{fig:BolModel} and~\ref{fig:BandModel}). We first consider in \S~ \ref{sec:early} the behavior at $0.8\,{\rm d}< t<6$~d, where the bolometric luminosity and effective temperature, and hence the photospheric radius, are well determined and their temporal evolution may be described by power-laws {(as explained in \S~\ref{sec:obs}, the data at earlier time do not allow for a reliable determination of temperature and bolometric luminosity, and are thus not used in constraining model parameters)}. We show that the luminosity and photospheric radius at $t\approx1$~d enable one to determine the ejecta mass $M$, the opacity $\kappa_{\rm M}$, and the energy deposition rate, $(1-f_{\nu\gamma})\dot{\varepsilon}_{\rm M}$. The inferred values of the parameters imply that the e$^\pm$ energy deposition becomes inefficient at $t\sim7$~d. This should lead to a steepening of the bolometric light curve, which is consistent with observations. The inferred parameter values also imply that the ejecta should at the same time become marginally optically thin. The similarity of the transparency time, $t_{\rm ph}$, and $t_\varepsilon$, when the e$^\pm$ energy deposition becomes inefficient, is not a coincidence. Rather, it is a results of the inferred low value of the opacity (see \S~\ref{sec:ModelDesc}).

The emission at $t>7$~d, when the ejecta becomes optically thin, is discussed in \S~\ref{sec:late}, with a focus on the constraints implied by the IR emission on the composition of the ejecta. We defer the discussion of the implications of the low opacity inferred in \S~\ref{sec:early} to this section. The results of our simple model are shown to be consistent with the results of detailed numerical calculations in \S~\ref{sec:comp_numerical}, where we also discuss the limitations of numerical calculations due to uncertainties in the opacities.

\subsection{The $t<6$~d emission}
\label{sec:early}

\begin{figure}
\hskip0.5cm
\includegraphics[width=8cm]{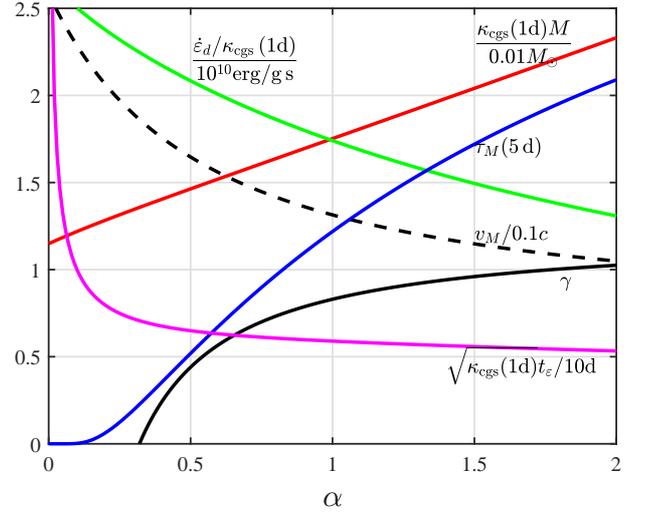}
\caption{Constraints implied by the observations on model parameters. The red, green and black dashed curves show the values of $\kappa_{\rm M} M$, $\dot{\varepsilon}_d/\kappa_{\rm M}$ and $v_{\rm M}$ respectively, inferred from the observed luminosity and photospheric radius at $t=1$~d for different values of $\alpha$, assuming $t_{\rm M}=1$~d and using Eqs.~(\ref{eq:tM_num}), (\ref{eq:LM_num}), and (\ref{eq:r_phot}) ($\kappa_{\rm cgs}(1d)=\kappa_{\rm M}/1{\rm cm^2g^{-1}}$). The solid black curve shows $\gamma(\alpha)$ inferred from the temporal evolution of $r_{\rm ph}$ (Eq.\ref{eq:r_phot}). The blue and magenta curves show the model prediction, based on the inferred parameter values, for the optical depth of the ejecta at $t=5$~d, $\tau_{\rm M}(5{\rm d})$, and for the time $t_\varepsilon$ at which the e$^\pm$ energy deposition becomes inefficient.
\label{fig:parameter_fit}}
\end{figure}

As we show below, the luminosity at $\sim1$~d requires the opacity to be low, implying that the diffusion front crosses the ejecta before the energy deposition of the electrons and positrons is suppressed, i.e. $t_{\rm M}<t_\varepsilon$ ($X_{\rm M}<1$, see Eq.~\ref{eq:XM}). The decline of the bolometric luminosity at later time, $L\propto t^{-1}$ up to $\sim6$~d (see Fig.~\ref{fig:LumBolFit}, table~\ref{tab:pow_fit}), may reflect in this case the decline of the radioactive energy release, $L\approx M\dot{\varepsilon}\propto t^{-\beta}$ with $\beta=1$, provided that the diffusion wave penetrates through the entire ejecta by $t=1$~d, i.e. $t_{\rm M}\le1$~d. Let us assume first that $t_{\rm M}\approx1$~d (as we explain below, this must be the case). The ejecta mass $M$ is determined in this case by Eq.~(\ref{eq:tM_num}), which gives $t_{\rm M}(M,\kappa_{\rm M},v_{\rm M})$, and by Eq.~(\ref{eq:r_phot}), which relates $v_{\rm M}$, the velocity of the slower part of the ejecta, to the observed photospheric velocity at $1$~d, $v_{\rm ph} = r_{\rm ph}(1{\rm d})/1{\rm d}=6.7\times10^{14}{\rm cm}/1{\rm d}=0.26c$ (see Fig.~\ref{fig:V}, Table~\ref{tab:BolLum}). The inferred mass depends weakly on the assumed value of $\alpha$, as shown in Fig.~\ref{fig:parameter_fit}, and is given by
\begin{equation}\label{eq:M}
  M \approx 1.5\times10^{-2} \left(\frac{\kappa_{\rm M}}{1{\rm cm^2/g}}\right)^{-1}M_\odot.
\end{equation}
Since $L_{\rm M}=(1-f_{\nu\gamma})\dot{\varepsilon}_{\rm M} M$, the observed luminosity at 1~d, $L=6.1\times10^{41}{\rm erg/s}$ (see Fig.~\ref{fig:LumBolFit}, Table~\ref{tab:pow_fit}), determines the ratio between the energy deposition rate and the opacity (see Eq.~\ref{eq:LM_num} and Fig.~\ref{fig:parameter_fit}),
\begin{equation}\label{eq:e-dotM}
  \dot{\varepsilon}_{\rm M}\approx 6\times10^{10}\frac{1}{3(1-f_{\nu\gamma})}\frac{\kappa_{\rm M}}{1{\rm cm^2/g}}{\rm\frac{erg}{g\,s}}.
\end{equation}
For a decay rate $\dot{\varepsilon}\propto t^{-1}$ over $\sim1$~s to $\sim 5$~d  the total radioactive decay energy is $\approx 10t_{\rm M}\dot{\varepsilon}_{\rm M}$, corresponding to a radioactive decay energy per nucleus of
 \begin{equation}\label{eq:eA-cons}
  E_A\approx 7\frac{1}{3(1-f_{\nu\gamma})}\frac{\kappa_{\rm M}}{1{\rm cm^2/g}}{\frac{A}{100}}{\rm MeV}.
\end{equation}
Here $A$ is the mass number of the nucleus (note that this is the average energy per all nuclei in the ejecta). This result implies that the opacity of the ejected material cannot be larger than $\sim1{\rm cm^2/g}$, since $\kappa\gg1{\rm cm^2/g}$ would require radioactive energy release per nucleus which is well above that expected from beta decay. For typical radioactive energy release rates obtained in numerical calculations at 1~d for $Y_{\rm e}<0.4$, $\approx 2\times10^{10}{\rm erg/s\,g}$ \citep[e.g.][]{Freiburghaus99,Metzger10,Goriely11,Roberts11,Korobkin12,Wanajo14,Lippuner15,Barneskasen16,Hotokezaka16heat,Rosswog17}, we find
\begin{equation}\label{eq:kappaM}
  \kappa_{\rm M}\approx0.3{\rm cm^2/g}.
\end{equation}
The typical radioactive energy release rate at 1~d is significantly lower than $10^{10}{\rm erg/s\,g}$ for $Y_{\rm e}> 0.4$, and hence inconsistent with the observations \citep[see also][]{Rosswog17}.

It is important to emphasize that the conclusion, that the luminosity is generated at $t\sim1$~d by low opacity material, is independent of model assumptions. The mass given is Eq.~(\ref{eq:M}) is the maximum mass that can contribute to $L$ at 1~d, since this is the mass through which the diffusion wave can penetrate for a given opacity. This in turn implies that the values of $\dot{\varepsilon}/\kappa_{\rm M}$ and $E_A/\kappa_{\rm M}$ given in  Eqs.~(\ref{eq:e-dotM}) and~(\ref{eq:eA-cons}) are model independent lower limits for these parameters, and hence that $\kappa_{\rm M}\lesssim0.3{\rm cm^2/g}$ is a model independent upper limit on the opacity. Since the opacity cannot be significantly lower than this value, we also conclude that $t_{\rm M}$ cannot be significantly smaller than $1$~d. For $t_{\rm M}\ll1$~d, $M$ would be significantly smaller than given by Eq.~(\ref{eq:M}), and the value of $\dot{\varepsilon}$ required for producing the observed luminosity would exceed the value that may be provided by $\beta$ decay.

The temporal dependence of $r_{\rm ph}$, $r_{\rm ph}\propto t^{0.6}$ (see Fig.~\ref{fig:R}), implies a relation between $\gamma$ and $\alpha$ through Eq.~(\ref{eq:r_phot}), $(1+\gamma\alpha)/(1+2\alpha)=0.6$ (see Fig.~\ref{fig:parameter_fit}). The extension of the power-law behavior to $\sim5$~d implies that the optical depth of the ejecta $\tau_{\rm M}$ ($\tau(m=1)$, see Eq.~\ref{eq:tau}) is larger than unity up to that time. Fig.~\ref{fig:parameter_fit} shows $\tau_{\rm M}(t=5{\rm d})$ as function of $\alpha$, for the mass $M(\alpha)$ and velocity $v_{\rm M}(\alpha)$ inferred from the observations. In order to satisfy $\tau_{\rm M}(5{\rm d})>1$, we require $\alpha\gtrsim 0.5$, which implies $0.5\lesssim \gamma\lesssim 1$. Finally, the data shows evidence for a flat dependence of $L$ on $t$ at early time, $L\propto t^{0}$ at $t<t_{\rm M}\approx1$~d. Adopting this as a constraint, and using Eq.~(\ref{eq:L_tM_tE2}), we find $\gamma\approx\alpha\approx 0.6$, consistent with the constraint $\alpha>0.5$ derived from the requirement of $\tau_{\rm M}(5{\rm d})>1$. This value of $\gamma$ implies that the opacity does not evolve strongly with time.

The inferred low value of $\kappa_{\rm M}$ implies that the ejecta should become marginally optically thin, and that the e$^\pm$ energy deposition should become inefficient, at similar times. For the ejecta parameters inferred above, we have $t_{\rm ph}\approx t_\varepsilon \approx 7$~d. At $t>t_\varepsilon$ we expect a steepening of $L(t)$ by an additional factor $(t/t_\varepsilon)^{-2}$, which is consistent with the observed light curve (see Fig.~\ref{fig:LumBolFit}, Table~\ref{tab:pow_fit}). The deviation of the spectral distribution from a blackbody distribution at $t\gtrsim 5$~d is also consistent with the expected transition to optically thin emission, see figures~\ref{fig:PhotSpectBB}-\ref{fig:LumBolFit}.

\begin{figure}
\centerline{\includegraphics[width=8cm]{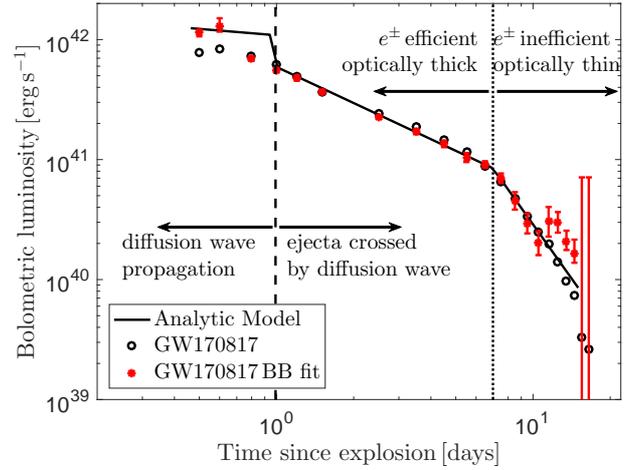}}
\caption{A comparison of the observed bolometric light curve {(black points and red stars)} and the analytic model light curve (solid black line), Eq.~(\ref{eq:L_tM_tE2}) with $t_{\rm M}=1$~d, $L_{\rm M}=6\times10^{41}{\rm erg/s}$, $T_{\rm M}=6600$~K, and $\{\alpha=0.7,\beta=1,\gamma=0.6,X_{\rm M}=0.02\}$ (the {black} points show the luminosity obtained from a photometric integration, and the {red stars} are obtained from the black-body fits to the data, see Fig.~\ref{fig:LumBolFit}). The values of $t_{\rm M}$, $L_{\rm M}$ and $T_{\rm M}$ are inferred directly from the observations, and determine the ejecta parameters $\kappa_{\rm M} M$, $\dot{\varepsilon}_{\rm M}/\kappa_{\rm M}$ and $v_{\rm M}$ (see Eqs.~\ref{eq:M} and~\ref{eq:e-dotM}, Fig.~\ref{fig:parameter_fit}). The large value of $\dot{\varepsilon}_{\rm M}/\kappa_{\rm M}$ requires a low value of $\kappa_{\rm M}$, $\kappa_{\rm M}\lesssim0.3{\rm cm^2/g}$. Since the value of $\kappa_{\rm M}$ cannot be much lower, this enables a determination of the ejecta parameters with little uncertainty. For the model shown in the figure, $M=0.05\,M_\odot$, $v_{\rm M}/c=0.15$, $\kappa_{\rm M}=0.3{\rm cm^2/g}$, $\kappa_e=0.4{\rm cm^2/g}$, and $(1-f_{\nu\gamma})\dot{\varepsilon}_{\rm M}=6\times10^9{\rm erg/s\,g}$. The model parameters are determined by the $t<6$~d observations. The low value of $\kappa_{\rm M}$ implies that the ejecta should become marginally optically thin, and that the e$^\pm$ deposition efficiency should drop below unity, at roughly the same time, $\sim7$~d (see Fig.~\ref{fig:parameter_fit}). This is consistent with the observed steepening of the luminosity decline, from $t^{-1}$ to $t^{-3}$, and with the increasing deviation of the observed spectra from thermal spectra, at $t>7$~d. We note that the analytic model parameters were not obtained as a best fit to the data. The model light curve is overlayed on the data to demonstrate that it captures both the qualitative and quantitative behavior. {The discontinuity at $t=t_{\rm M}$ is due to the simplified analytic treatment (see next to last paragraph of \S~\ref{sec:escape}), and is expected to be a smooth transition in reality.}
\label{fig:BolModel}}
\end{figure}

\begin{figure}
\centerline{\includegraphics[width=8cm]{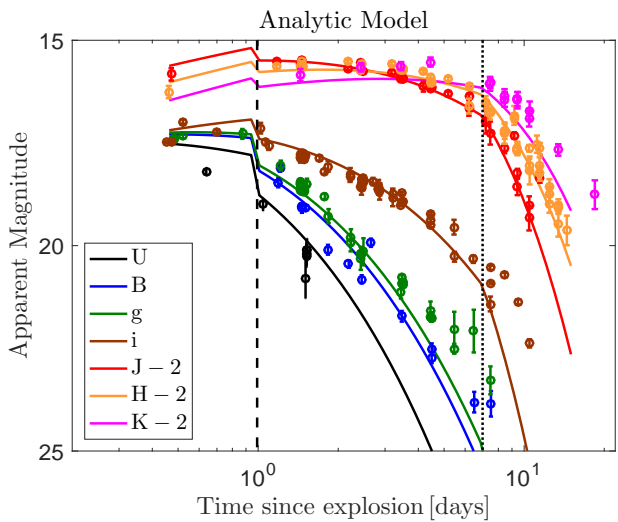}}
\caption{A comparison of the observed multi-band light curves and the analytic model multi-band light curves, given by Eqs.~(\ref{eq:L_tM_tE2}), (\ref{eq:r_phot}) and~(\ref{eq:TM_num}), for the same model parameters used in Fig.~\ref{fig:BolModel} (and including extinction). The model's spectral energy distributions are obtained assuming a black-body spectrum. This assumption breaks at $t>7$~d (the dotted line marks the time at which the ejecta becomes marginally optically thin, see Fig.~\ref{fig:parameter_fit}). Thus, while we expect the model to provide a good estimate of the bolometric luminosity at $t>7$~d, we do not expect it to provide an accurate description of the spectra. The fact that the observed spectrum is not far from thermal at $t\sim7$~d implies that the opacity at this time is $\sim0.1{\rm cm^2/g}$ at $1-2\mu$m (see discussion in \S~\ref{sec:late}). An accurate description of the spectrum at late times, when the ejecta becomes optically thin, requires an accurate knowledge of the non-LTE spectral opacity and is beyond the scope of the current analytic model (as well as of the scope of current detailed numerical calculations).
\label{fig:BandModel}}
\end{figure}

Figures~\ref{fig:BolModel} and~\ref{fig:BandModel} present a comparison of the observations with the bolometric and multi-band light-curves of the simple analytic model (for model parameter values as determined above). The simple model provides a good description of both the qualitative and quantitative properties of the observations.

For both Iron group elements and Lanthanides, a rapid decline of the opacity is obtained below $\sim3000$~K, due to recombination. Since the ejecta temperature drops to $\sim3000$~K at $t\sim 7$d, it may become optically thin at later time also for masses larger than $M$ inferred above. Thus, while the transition to optically thin emission expected for the above inferred parameters is consistent with observations, we do not consider this as strong evidence in support of the model. We also note, that the rapid drop in opacity implies that the temperature of the plasma cannot decreases significantly below 3000~K, as otherwise the emissivity would not be sufficient to enable the plasma to radiate the deposited energy.

As demonstrated in Fig.~\ref{fig:LumBolFit}, the bolometric light curve at $t>3$~d is consistent with an exponential decline with a time constant of $t_0\approx3.7$~d. This may suggest that an early power-law decay of the radioactive energy generated by the decay of a wide range of nuclei is overtaken at late time by an exponential dominated by a single isotope. Such an interpretation of the data would require $t_\varepsilon>10$~d, which implies $\kappa_e/\kappa_{\rm M}\gtrsim4$, since otherwise the luminosity decline would be faster than the intrinsic exponential due to the inefficiency of the e$^\pm$ energy deposition, i.e. at $t>t_\varepsilon$ we have $L\propto (t/t_\varepsilon)^{-2}\exp(-t/t_0)$. This interpretation of the data would however be inconsistent with the temporal dependence of the radioactive energy release obtained in detailed calculations.

\subsection{The behavior at late time, $t>6$~d}
\label{sec:late}

Based on the above analysis, we expect the ejecta to become marginally optically thin between 5 and 7~d (see Fig.~\ref{fig:parameter_fit}). At later time, the spectral luminosity, $L_\nu\equiv dL/d\nu$, is given by $L_\nu=\kappa_\nu M c u^{BB}_\nu(T)$, where $\kappa_\nu$ is the opacity and $u^{BB}_\nu(T)$ is the blackbody spectral energy density at the plasma temperature $T$. At the transition to optically thin emission, $t\sim6$\,d, the spectral luminosity is close to the blackbody luminosity, $L_\nu=4\pi R^2(c/4) u^{BB}_\nu(T)$, over the observed wavelength range, 0.5 to 2.5\,$\mu$m (see Figs.~\ref{fig:SpecSpectBB},~\ref{fig:PhotSpectBB}). This implies that the optical depth throughout this range is not far below unity at this time, i.e. that $\kappa_\nu M/\pi R^2\gtrsim1$. For $M\approx 0.05$\,M$_\odot$ and $v_{\rm M}/c\approx 0.1$, as inferred above, this implies $\kappa_\nu\gtrsim0.1$\,cm$^{2}$\,g$^{-1}$. Note that at later time, the decrease of $L$, $L\propto L^{-3}$, implies that the ratio of the spectral luminosity to the blackbody luminosity drops as $t^{-5}$ (since the radius increases like $t$ and the plasma temperature does not decrease significantly below 0.3\,eV).

\begin{figure}
\centerline{\includegraphics[width=8cm]{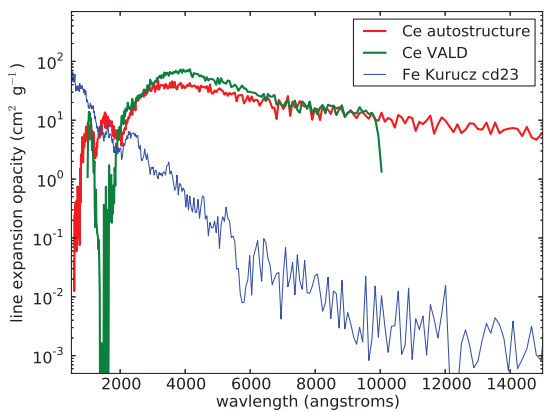}}
\caption{A comparison of the line expansion opacity for Fe and Ce, at $T=5000$\,K, $\rho=10^{-13}$\,g\,cm$^{-3}$ and $t=1$\,d, adopted from \citet{Kasen13}, using the Autostructure, Kurucz and VALD line lists (Note that since the expansion opacity is proportional to $1/(\rho c t)$, the implied opacity for our inferred model parameters at $t=7$\,d would be $\approx 7$ times larger than shown in this plot).
\label{fig:Fe_vs_Ce_Kurucz}}
\end{figure}

The low value of the frequency averaged (expansion) opacity, $\kappa_{\rm M}\approx0.3$\,cm$^{2}$\,g$^{-1}$, required to account for the bolometric light curve and for the photospheric radius, may be provided for example by Iron group elements, and requires no contribution from high opacity elements, such as the Lanthanides. The requirement that the spectral (expansion) opacity $\kappa_\nu$ maintains a value $\approx 0.1{\rm cm^2/g}$ through the 1-2~$\mu$m range yields additional constraints. The calculations of \citet{Kasen13} and \citet{BarnesKasen13}, based on the Kurucz/Autostructure line lists, suggest that the opacity of Iron group elements drops with wavelength to values well below $0.1{\rm cm^2/g}$, $\kappa_\nu<10^{-2}$\,cm$^{2}$\,g$^{-1}$ at $\lambda\gtrsim1$\,$\mu$m, while the opacity of the Lanthanides is much higher at IR wavelengths, $\kappa_\nu\approx10^{2}$\,cm$^{2}$\,g$^{-1}$, see Fig.~\ref{fig:Fe_vs_Ce_Kurucz}. This implies that Iron group elements will not be able to produce the observed IR luminosity, while the Lanthanides may. The large Lanthanides' opacity implies that a small mass fraction of these elements, $\approx 10^{-3}$, would be sufficient to produce the required opacity. Such a small mass fraction would not affect significantly the frequency averaged (expansion) opacity, i.e. would not increase $\kappa_{\rm M}$ beyond $\approx0.3$\,cm$^{2}$\,g$^{-1}$, and is therefore consistent with the constraints derived from the earlier time behavior. A significantly larger Lanthanides' mass fraction will lead to a significant increase of the averaged opacity, $\kappa_{\rm M}$, which will suppress the emission at early time. A larger mass fraction is therefore excluded in a single component ejecta model.

The opacities of the Lanthanide elements are not well known. The line data available are limited, and the theoretical calculations are highly complex. As a result, the opacity of the r-process mixture of Lanthanide elements is estimated using theoretical model calculations of a few elements, which are considered representative (Nd, Ce). Different approximations lead to large variations in the derived opacity of individual elements \citep[see, e.g., Fig. 8 of][]{Kasen13}, and the uncertainty in the opacity of the mixture exceeds an order of magnitude \citep[see e.g. Figure 1 of][]{BarnesKasen13}. This implies that the mass fraction of Lanthanides required to provide the inferred opacity is uncertain.

\begin{figure}
\centerline{\includegraphics[width=8cm]{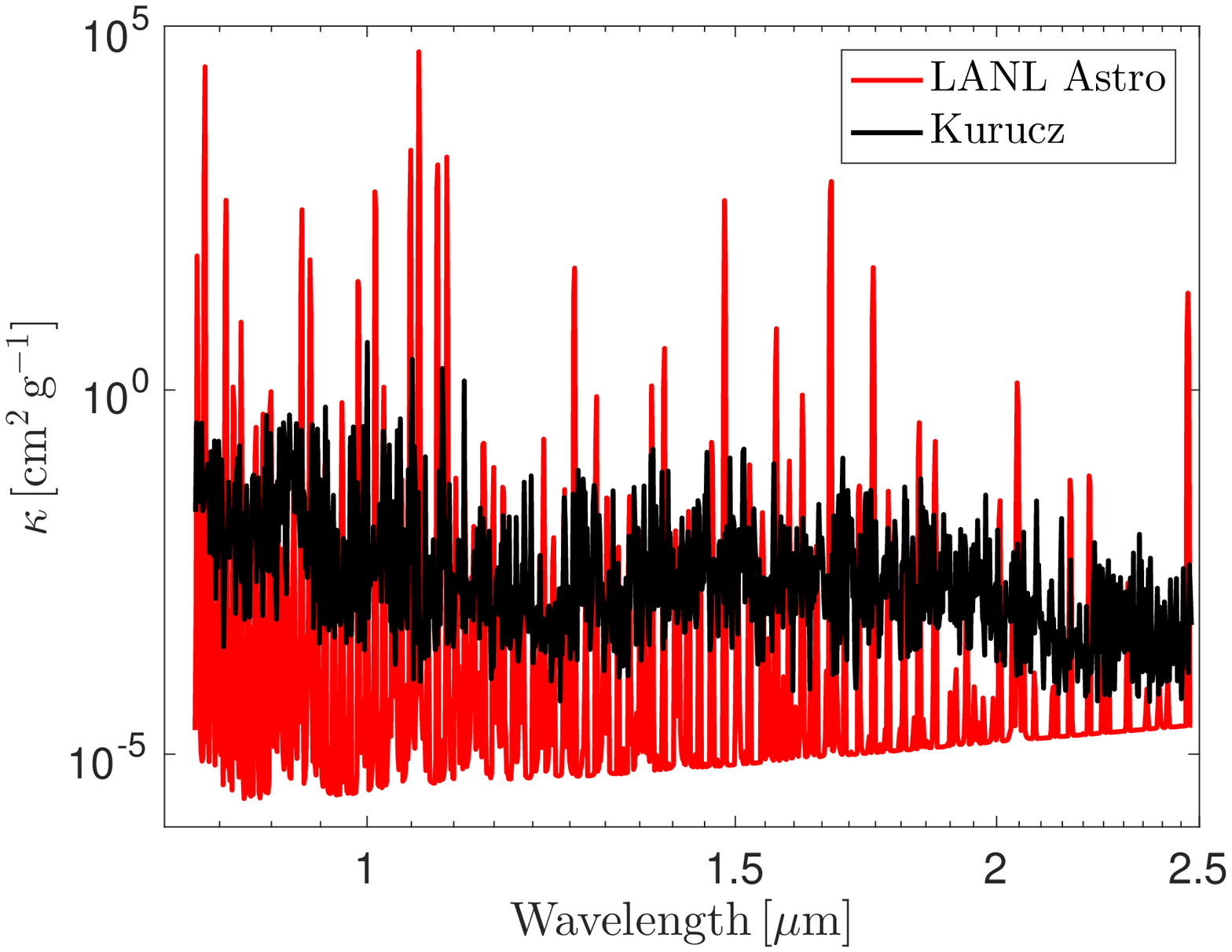}}
\caption{A comparison of the Planck opacity for Fe, at $T=0.5$\,eV and $\rho=10^{-13}$\,g\,cm$^{-3}$, obtained by the LANL Astrophysical Opacity calculator (\texttt{http://aphysics2.lanl.gov/cgi-bin/opacrun/astro.pl}, $d\lambda/\lambda=10^{-3}$, {red}) and using the Kurucz line list \citep[][{, black}]{Kurucz93,Kurucz95}. 
The LANL calculation yields multiple lines in the IR range with oscillator strengths that exceed by some 6 orders of magnitude those obtained in the calculation based on the Kurucz line list.
\label{fig:LANL_Kurucz}}
\end{figure}

Large uncertainties exist also in the opacity of lower mass elements. Figure~\ref{fig:LANL_Kurucz} compares the opacities derived for Iron at 0.5~eV using the Kurucz line list with those obtained from the LANL Astrophysical Opacity calculator. As the plot clearly demonstrates, the LANL calculation yields multiple lines in the IR range with oscillator strengths that exceed by some six orders of magnitude those obtained in the calculation based on the Kurucz line list. The expansion opacity that would be produced by such lines would be more than sufficient to yield large optical depth, and near blackbody emission, in the IR. The LANL calculations are limited to temperatures $T\ge0.5$~eV, and the lines shown at 0.5~eV may not be relevant to the lower temperature, $T\simeq0.3$~eV in which we are interested. However, the very large discrepancy between the LANL and Kurucz line list based calculations implies that careful examination of the line opacities is necessary in order to draw robust conclusions regarding the composition of the ejecta.

{It should be noted here, that constraints on the opacity of Iron at $1-2\mu$m, at the conditions relevant for NS merger ejecta at the relevant time of $t\gtrsim 5$~d, cannot be directly obtained from observations of Type Ia supernovae due to several reasons. First, due to the larger mass ($M\sim1 M_\odot$) and smaller velocity ($v\sim c/30$) characterizing the ejecta of type Ia SNe, a given column density, $M/(4\pi r^2)=M/(4\pi v^2 t^2)$, is reached by the Ia ejecta at a much later time than that at which it is reached by the NS merger ejecta. In particular, a column density of $1{\rm g/cm}^2$ is reached by the NS merger ejecta at $\approx 7$~d, and by the SN Ia ejecta at $\approx 150$~d. Thus, only IR observations at very late times, $t>150$~d, probe the relevant column densities (and hence optical depths). Second, since the expansion opacity is proportional to the velocity gradient, the larger characteristic velocity of the NS merger ejecta implies that the effective opacity of the NS merger ejecta may be significantly larger than that of the Ia ejecta under similar conditions (temperature, density, column density). Finally, we are interested in the opacity at the time at which the optical depth drops below unity. At this time, the opacity is not given by its LTE value, and depends strongly on the density of the radiation field. Since the luminosity of Ia SNe at $\sim150$~d is similar to that of the NS merger ejecta at $\sim 7$~d, while their radii are $\approx 4$ times larger ($r\propto M^{1/2}$ for a given column density), the radiation density is $\sim100$ times lower in the Ia ejecta compared to the NS merger ejecta (when the two reach the same column density).}

Adopting the \citet{Kasen13} derived opacities, our analysis implies that the observations are consistent with a single composition ejecta, with Lanthanides' mass fraction of $\approx 10^{-3}$. One may consider the possibility of the existence of another ejecta component, at $v\lesssim v_{\rm M}$, with Lanthanides' mass fraction $\gg 10^{-3}$. If such a shell is contributing significantly to the luminosity at $t\ge 7$~d, then its mass is limited by the rapid decrease of $L$, $L\propto t^{-3}$, at this time. This rapid decrease requires the energy deposition by e$^\pm$ to be inefficient, $(c/v)\kappa_e\rho R<1$, and the photon escape time to be shorter than $t$, $\kappa\rho R<c/v$ (As noted in\S~\ref{sec:early}, a rapid decrease in $L$ may also be obtained by a rapid, exponential, decrease in the radioactive energy release rate. However, such an exponential decay is not expected for Lanthanides rich material originating from a low $Y_{\rm e}$ ejecta). The resulting limit on $\rho R$, $\rho R<\min[(v/c)\kappa_e^{-1}, (c/v)\kappa^{-1}]<0.1(v/0.1c)$\,g\,cm$^{-2}$, sets an upper limit to the shell's mass, $\lesssim2\times10^{-3}$M$_\odot$ for $v/c=0.1$. We cannot, of course, rule our the existence of a slower shell with larger mass and opacity, that does not contribute significantly to the observed luminosity.

\subsection{Comparison to detailed numerical calculations}
\label{sec:comp_numerical}

\begin{figure}
\centerline{\includegraphics[width=8cm]{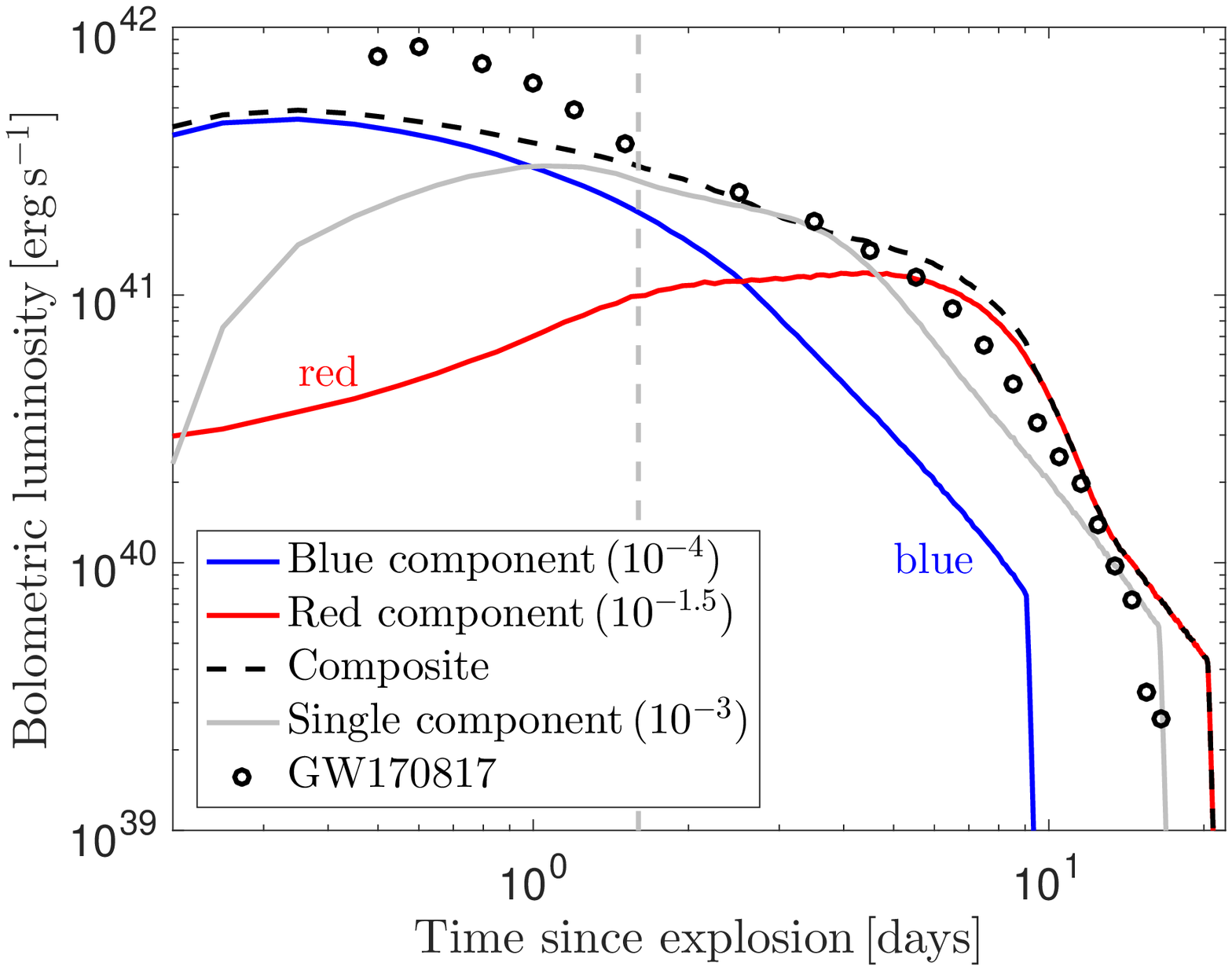}}
\caption{A comparison of the observed bolometric light curve ({black}) with those obtained in the detailed numerical calculations of \citet{Kasen17Nat} (available at \texttt{https://github.com/dnkasen/Kasen\_Kilonova\_{\rm M}odels\_2017}), for different values of the Lanthanides' mass fraction (as shown in the inset). The ejecta mass and velocity are $\{0.025$M$_\odot,0.3c\}$, $\{0.04$M$_\odot,0.15c\}$, and $\{0.04$M$_\odot,0.2c\}$ for the mass fractions of $10^{-4}$ (blue), $10^{-1.5}$ (red) and $10^{-3}$ ({gray}) respectively. The black-dashed curve is the 2-component model of \citet{Kasen17Nat}, obtained by summing the luminosities obtained for the $10^{-4}$ and $10^{-1.5}$ mass fraction calculations. The dashed {gray} line shows the time at which the diffusion sphere crosses most of the ejecta, $t_{\rm M}$, in the single component model. For similar characteristic ejecta parameters $\{M,v,\kappa\}$, $t_{\rm M}$ is larger in the numerical models than in the analytic model due to the wider velocity distribution of the analytic model, with larger mass at high velocity (see text). This is likely the reason for the lower flux predicted by the numerical models (compared to the the analytic model) at early times.
\label{fig:Kasen_Bol}}
\end{figure}

\begin{figure}
\centerline{\includegraphics[width=8cm]{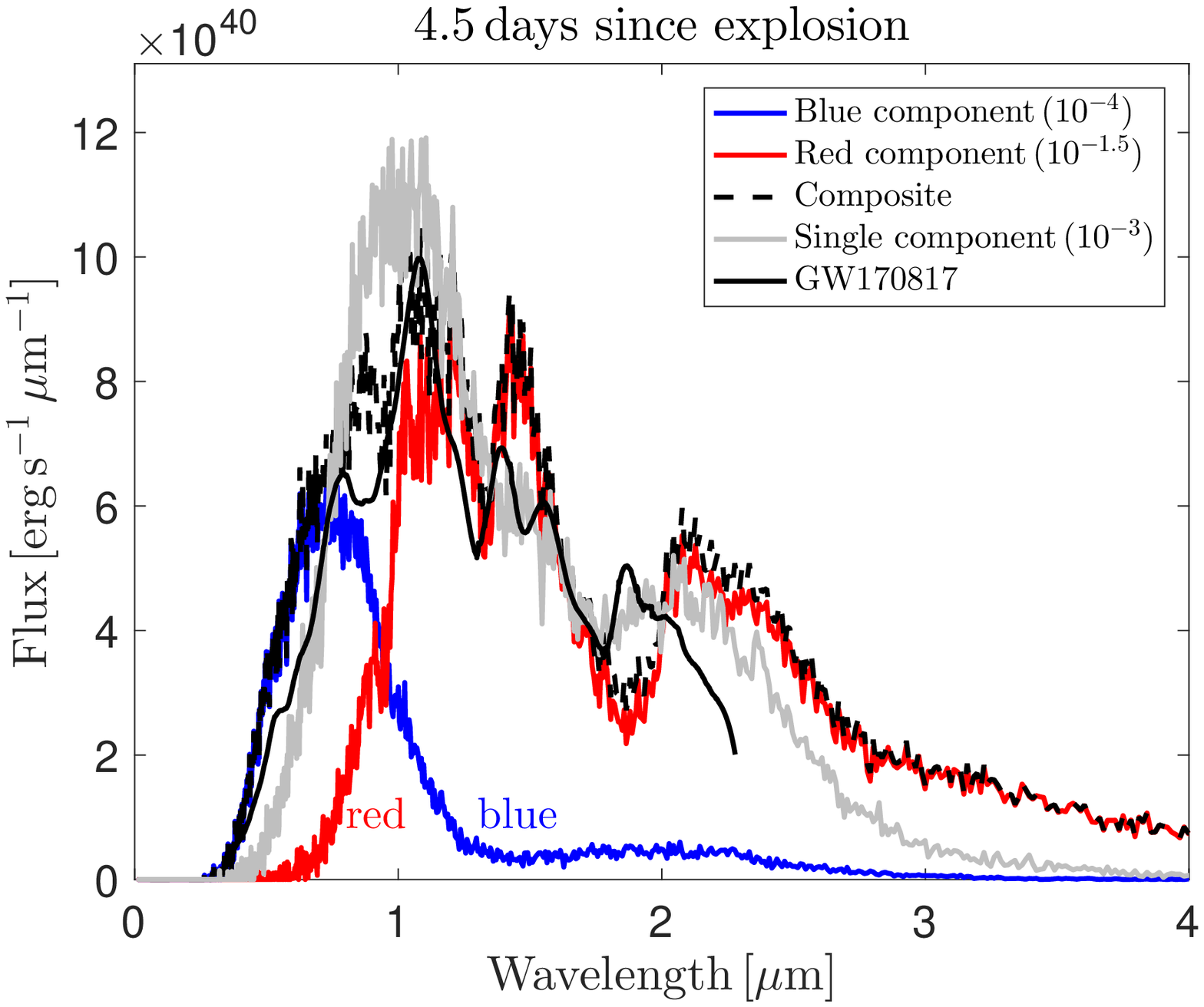}}
\caption{A comparison of the observed spectral luminosity ({black}) with those obtained in the detailed numerical calculations of \citet{Kasen17Nat} (avilable at \texttt{https://github.com/dnkasen/Kasen\_Kilonova\_{\rm M}odels\_2017}) at $t=4.5$\,d, for different values of the Lanthanides' mass fraction. The ejecta mass and velocity are $\{0.025$M$_\odot,0.3c\}$, $\{0.04$M$_\odot,0.15c\}$, and $\{0.04$M$_\odot,0.2c\}$ for the mass fractions of $10^{-4}$ (blue), $10^{-1.5}$ (red) and $10^{-3}$ ({gray}) respectively. The black-dashed curve is the 2-component model of \citet{Kasen17Nat}, obtained by summing the spectral luminosities obtained for the $10^{-4}$ and $10^{-1.5}$ mass fraction calculations. {The plots confirm our conclusion that a $10^{-3}$ mass fraction of Lanthanides would produce an opacity in the 1--2$\mu$ band that is sufficient to account for the observed IR luminosity.}
\label{fig:KasenSpec1}}
\end{figure}
\begin{figure}
\centerline{\includegraphics[width=8cm]{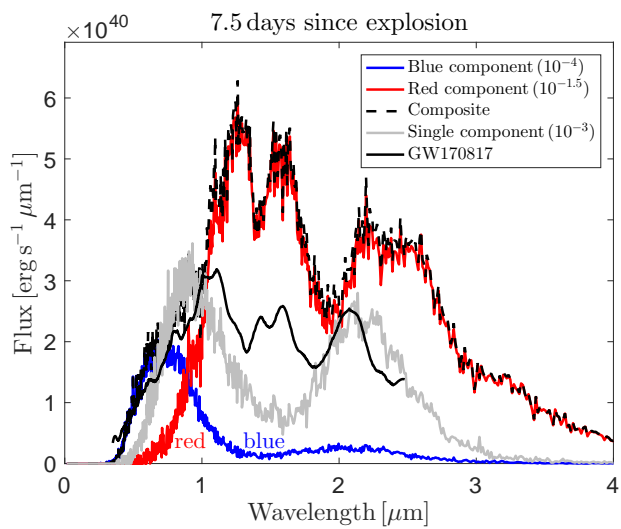}}
\caption{Same as Figure~\ref{fig:KasenSpec1}, for $t=7.5$\,d.
\label{fig:KasenSpec2}}
\end{figure}

We compare in this section our results to those obtained by the detailed numerical calculations of \citet{Kasen17Nat}, which are publicly available. A direct comparison of the analytic model to the numerical calculations is not possible, due to the different velocity distributions used. In the numeric calculations, most of the mass (of a single ejecta component) is concentrated around a single characteristic velocity $v_t$, with $dm/dv\propto v$ for $v<v_t$ and $dm/dv\propto v^{-8}$ for $v>v_t$. Our analytic model has a wider velocity distribution, $dm/dv\propto v^{-(\alpha+1)/\alpha}\approx v^{-2.5}$, with a significant fraction of the mass at large velocity \citep[our velocity distribution is wider also than that used in the semi-analytic models of][where $dm/dv\propto v^{-4}$]{Metzger17rev,2017arXiv171011576V,CowperthwaiteBerger173KN}. This implies that the temporal dependence of the photospheric temperature (radius) obtained in the numerical calculations differs from that of the analytic model, and hence also from the observed dependence. Moreover, the bolometric luminosity of the numeric model would differ from that of the analytic model up to the time at which the diffusion sphere crosses most of the ejecta (after which $L=M\dot{\varepsilon}_d$). Finally, if the adiabatic energy loss of the e$^\pm$ is neglected in the numerical calculations, the bolometric luminosity of the numeric models would exceed that of the analytic model at late times.

Given the above mentioned limitations, the main goal of the comparison to numerical results is to test our conclusion, that the late time IR emission may be produced by an ejecta with $\kappa_\nu\approx0.1{\rm cm^2/g}$ at $1-2\mu$m, corresponding to a small, $\approx10^{-3}$, mass fraction of Lanthanides (adopting the opacities used in the numerical calculations).

Figure~\ref{fig:Kasen_Bol} compares the observed bolometric light curve with those obtained in the detailed numerical calculations of \citet{Kasen17Nat} for different values of Lanthanides' mass fraction. The single component ejecta, with Lanthanides' mass fraction of $\approx 10^{-3}$, provides as good a fit to the data as provided by the two component ejecta model. Figures~\ref{fig:KasenSpec1} and~\ref{fig:KasenSpec2} compare the observed spectra at 4.5~d and 7.5~d with those predicted by the calculations. The single ejecta model produces a reasonable fit to the observations at both times, confirming our conclusion that a $10^{-3}$ mass fraction of Lanthanides would produce an opacity that is sufficient to account for the observed IR luminosity.

A note is in place here regarding the low luminosity produced by the numerical models compared to the analytic model and to the observations {\citep[this has led to the suggestions of additional sources of luminosity at early times, e.g.][]{Piro17,MetzgerThompson08}.} Due to the different velocity profiles of the numeric and analytic models, for similar characteristic ejecta parameters, $\{M,v_t=v_{\rm M},\kappa\}$, the time at which the diffusion sphere penetrates through the ejecta is longer in the numerical model (by a factor of $\sim2$) compared to that obtained in the analytic model. This is likely the reason for the lower flux predicted by the numerical models (compared to the analytic model) at early times. Due to the narrower velocity distribution, the time at which the e$^\pm$ energy deposition becomes inefficient is also longer in the numerical models than in the analytic model. For the single component model shown in figure~\ref{fig:Kasen_Bol}, $t_\varepsilon\simeq13$~d.

As explained in the preceding section, the presence of a high opacity shell with significant contribution to the luminosity would be difficult to reconcile with the rapid drop of $L$ at late time. This is demonstrated in Fig.~\ref{fig:KasenSpec2}, where the two component model flux is shown to largely exceed the observed flux at 7.5~d. We do not consider this discrepancy as severe, since the model luminosity may be modified by modifying the energy deposition rate, which is not accurately known. However, a 2-component model would in any case require some tuning of the two components to produce the observed smooth evolution.

None of the numerical models produce a good description of the spectral minima and maxima. Here too, we do not consider this to be a severe discrepancy since the opacities are not accurately known. As discussed in detail in the preceding section, the opacities of individual Lanthanide elements are uncertain, and furthermore the opacity of the r-process mixture of Lanthanide elements is estimated using theoretical model calculations of a few elements, which are considered representative. This implies that the results of detailed calculations of spectra cannot be directly compared to the detailed structure of the observed spectra, as the numerical results depend strongly on the assumed opacity structure. As clearly demonstrated, e.g., in Figure 13 of \citet{Kasen13}, different approximations used in the opacity calculations lead to large variations in the predicted spectra. Finally, we note that the opacity uncertainties are even larger at late times, when the plasma becomes optically thin and deviates from LTE.

\section{Discussion}
\label{sec:discussion}

A simple model describing the emission from merger ejecta was presented in \S~\ref{sec:Model}. The main model results, including the predicted light curves, $L(t)$, photospheric radii, $r_{\rm ph}(t)$, and effective temperatures, $T_{\rm eff}(t)$, are given in \S~\ref{sec:ModelSum}. The light curve behavior is qualitatively different for low, $\kappa\lesssim1{\rm cm^2/g}$, and high, $\kappa\gg1{\rm cm^2/g}$, opacity. In particular, for low opacity the time at which the ejecta becomes optically thin, $t_{\rm ph}$, is similar to the time at which the e$^\pm$ energy deposition becomes inefficient, $t_{\rm ph}\sim t_\varepsilon$, while for large opacity $t_\varepsilon\ll t_{\rm ph}$.

We have shown in \S~\ref{sec:comp_obs} that the observed UV to IR radiation may be well reproduced by a single component ejecta, described by the simple analytic model of \S~\ref{sec:Model}. See figures~\ref{fig:BolModel} and~\ref{fig:BandModel}. The observations at $t<6$~d enable one to infer the ejecta parameters, $M\approx0.05~M_\odot$, $v_{\rm M}\approx 0.15c$, $\kappa_{\rm M}\approx 0.3$\,cm$^{2}$\,g$^{-1}$, $\dot{\varepsilon}\approx 2\times10^{10}(t/1~{\rm d})^{-1}{\rm erg\,g^{-1}s^{-1}}$, and $v\propto m^{-\alpha}$ with $\alpha\approx 0.7$ extending to $>0.3\,c$.  The implied energy generation rate is consistent with that predicted for a composition produced by the expansion of a $Y_{\rm e}<0.4$ ejecta. For the inferred ejecta parameters, the simple model predicts $t_{\rm ph}\approx t_\varepsilon \approx 7$\,d. This is consistent with the  steepening of $L(t)$ by a factor $(t/t_\varepsilon)^{-2}$ at $t>7$\,d (see Fig.~\ref{fig:LumBolFit}, Table~\ref{tab:pow_fit}), and with the deviation of the spectral distribution from a blackbody distribution at $t\gtrsim 5$\,d. {In our model, a low opacity, $0.3\le \kappa/(1{\rm cm^2/g})\le1$, ejecta accounts for the electromagnetic emission at all times. This is in contrast with multi-component models, in which the early, $t<2$~d, emission is dominated by low opacity, $\kappa\approx0.5{\rm cm^2/g}$, material, and the late, $t>5$~d, emission is dominated by high opacity, $\kappa\approx10{\rm cm^2/g}$, material \citep[e.g.][]{Kasen17Nat,2017arXiv171011576V}.}

The low frequency averaged (expansion) opacity, $\kappa_{\rm M}\approx0.3$\,cm$^{2}$\,g$^{-1}$, which is required to account for the bolometric light curve and for the inferred photospheric radii, may be provided for example by Iron group elements, and requires no contribution from high opacity elements, such as the Lanthanides. As explained in \S~\ref{sec:late}, the fact that the spectral luminosity at $t\sim7$\,d is close to the blackbody luminosity all the way up to $\lambda\sim2~\mu$m, requires that the spectral (expansion) opacity $\kappa_\nu$ maintains a value $\approx 0.1$\,cm$^{2}$\,g$^{-1}$ through the 1-2\,$\mu$m range. This requirement may imply more stringent constraints on the composition of the ejecta. If such opacity in the IR regime can only be provided by Lanthanides, as suggested by the calculations of \cite{Kasen13} based on the Kurucz/Autostructure line lists, then the mass fraction of these elements should be $\approx 10^{-3}$ (this mass fraction is sufficient to produce the required opacity, while a larger mass fraction would suppress the emission at early time and is therefore excluded in a single component ejecta model).

As discussed in some detail in \S~\ref{sec:late}, the opacities of the Lanthanide elements are not well known, which implies that the mass fraction of Lanthanides required to provided the inferred opacity is uncertain. Furthermore, we have shown in \S~\ref{sec:late} that large discrepancies exit between the opacities derived for Iron at 0.5~eV using the Kurucz line list and those obtained from the LANL Astrophysical Opacity calculator, see Figure~\ref{fig:LANL_Kurucz}. In particular, the LANL calculation yields multiple lines in the IR range with oscillator strengths that far exceed those obtained in the calculation based on the Kurucz line list. The large discrepancy implies that a careful examination of the line opacities is necessary in order to draw robust conclusions regarding the composition of the ejecta, and in particular regarding the required presence of the Lanthanides. It is important to note here that the uncertainties in the opacity are much larger still for non LTE plasma. Hence, a detailed description of the spectra at late time, when the ejecta becomes optically thin, is beyond the scope of both the current simple analytic model and the detailed numerical calculations.

A comparison of our results to those of detailed numeric calculations is given in \S~\ref{sec:comp_numerical}. A direct comparison is complicated, due to the different velocity distributions used. In the numeric calculations most of the mass (of a single ejecta component) is concentrated around a single characteristic velocity, while the velocity distribution of the analytic model is wider, $dm/dv\propto v^{-(\alpha+1)/\alpha}\approx v^{-2.5}$, with a significant fraction of the mass at large velocity. This implies that the temporal dependence of the photospheric temperature (radius), as well as the early evolution of the bolometric luminosity, obtained in the numerical calculations differ from those of the analytic model, and hence also from the observations (which are well described by the analytic model). The main goal of the comparison to numerical results is to test our conclusion, that the late time IR emission may be produced by an ejecta with $\kappa_\nu\approx0.1{\rm cm^2/g}$ at $1-2\mu$m, corresponding to a small, $\approx10^{-3}$, mass fraction of Lanthanides (adopting the opacities used in the numerical calculations). As can be seen in Figs.~\ref{fig:Kasen_Bol}-\ref{fig:KasenSpec2}, this conclusion is indeed supported by the detailed numerical calculations. The single component model with a $\approx 10^{-3}$ Lanthanides' mass fraction provides a good fit to the late time IR data, as well as a fit to the earlier time data, which is as good as that provided by the two component (mass fractions of $\approx 10^{-1.5}$ and $\approx 10^{-4}$) model.

Models with several ejecta components, dynamical, wind and secular, with large opacity differences predict a strong dependence on observing angle of the of the observed UV-IR signal. A single component model predicts only a weak dependence. This provides a clear test that may be used to discriminate between the models using future observations.

Most of the mass of r-process elements with solar abundance is concentrated in elements of mass number $70<A<90$ \citep{Sneden08R,Thielemann11R}. For $Y_{\rm e}\approx 0.4$, most of the ejecta mass is expected to be carried by elements in this mass range \citep[e.g.][]{Korobkin12,Rosswog17}. In this case, NS mergers may produce the observed solar abundance of these elements, as may be concluded from the following argument. For solar abundance the mass of $70<A<90$ elements is $\approx10^{-3}$ that of Iron. Since a significant fraction of this Iron mass may be produced in SN Ia, which occur at a rate $\sim10^{-4}{\rm Mpc^{-3}yr^{-1}}$ and release $\sim0.5M_\odot$ of Iron, the mass of the r-process elements may be provided by NS mergers that occur at a rate slower by a factor 100 and emit 10 times less mass.

A Lanthanides' mass fraction of $\approx 10^{-3}$ is well below the mass fraction of Lanthanides out of all $A\gtrsim70$ r-process nuclei in a solar composition material, $\approx3\%$ \citep{Sneden08R,Thielemann11R}. Moreover, as explained in \S~\ref{sec:late}, the rapid decrease of $L$ at $t>7$\,d sets an upper limit of $\approx2\times10^{-3}M_\odot$ on the mass of a second component of the ejecta, with $v\approx 0.1c\lesssim v_{\rm M}$ and a Lanthanides' mass fraction $\gg 10^{-3}$, that contributes significantly to the luminosity. Thus, while it is impossible to rule out the existence of a slow, Lanthanides rich component of the ejecta, that does not contribute to the observed luminosity, we do not have direct evidence for the production of "heavy r-process" elements with a mass fraction that would be required to account for the solar abundance of such elements.

The observed properties of the ejecta, $\sim 10^{-1.5}$M$_\odot$ with relatively large velocity, $0.15c$ to $0.3c$, and large value of $Y_{\rm e}$, that would be required to avoid a large mass fraction of Lanthanides, are inconsistent with the general predictions of numerical simulations. These generally predict that the fast component of the ejecta is produced by "dynamical" ejection, with little mass and low $Y_{\rm e}$ value, while the larger fraction of ejected mass is contributed by the "secular" ejection, which may be characterized by larger $Y_{\rm e}$ values but expected to be slow, $v<0.05c$. This difficulty has been recently pointed out also by \citet{Fujibayashi17}, who suggest that viscous heating due to magneto-hydrodynamic turbulence in the merger remnant may lead to the ejection of $\sim10^{-2}M_\odot$ with $Y_{\rm e}\gtrsim0.25$. It is, however, challenging to obtain the required high velocity also in this scenario. If the opacity is provided by Lanthanides, their required mass fraction, $\approx10^{-3}$, may be obtained for initial $Y_{\rm e}\approx 0.25$ \citep[e.g.][]{Korobkin12}, or by mixing of ejecta components with different mass fractions of Lanthanides, i.e. different initial values of $Y_{\rm e}$. The latter possibility may be more likely, since the mass fraction of Lanthanides varies strongly with $Y_{\rm e}$ at $Y_{\rm e}\approx 0.25$.

The prompt detection of GW\,170817 in the blue/UV bands has several implications for future searches for electromagnetic counterparts of NS merger events. At early times, NS mergers appear blue and have significant emission in the near UV (e.g., \citealt{2017arXiv171005437E}). It is likely that at earlier times ($<10$\,hr) the UV emission will dominate simply because the emission temperature would be higher
and any absorption features in the UV will be less prominent (as the gas is ionized). Therefore, future UV missions such as {\it ULTRASAT} \citep{2014AJ....147...79S}, that combines a large instantaneous field-of-view
($\approx245$\,deg$^{2}$) with significant depth (5-$\sigma$ limiting AB magnitude $\approx22.4$ in 900\,s),
and rapid target of opportunity capabilities ($<10$\,min) for a large fraction of the celestial
sphere ($\approx50$\% at any given moment) will be extremely effective electromagnetic counterpart survey machines.

\begin{figure}
\centerline{\includegraphics[width=8cm]{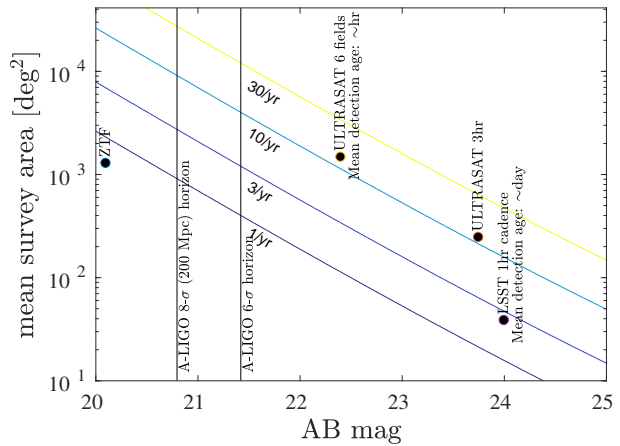}}
\caption{The expected detection rate of electromagnetic emission from NS-merger events not triggered by GW detection, as a function of survey limiting magnitude and mean observed area
(within the appropriate cadence).
Circles represent the capabilities of some surveys, under assumptions
specified in the text, while the vertical lines show
the LIGO horizon.
We assumed that all events will have peak UV to blue absolute magnitude
of about $-15.7$.
{This Figure does not provide a 1 to 1 comparison between surveys, rather it
demonstrates that with appropriate strategies future surveys may be able to detect
NS mergers without the GW signal.}
\label{fig:FutureDetectors}}
\end{figure}

Furthermore, it is possible that future missions like {\it ULTRASAT} and the Large Synoptic Survey Telescope \citep[LSST,][]{2008arXiv0805.2366I} will be able to detect NS merger events blindly, without a gravitational-wave detector trigger (e.g., beyond the LIGO horizon).
Figure~\ref{fig:FutureDetectors} shows the expected NS-merger electromagnetic counterpart detections
per year as a function of survey limiting magnitude and mean observed area
(within the appropriate cadence).
We assumed a NS-merger event rate of 1500\,Gpc$^{-3}$\,yr$^{-1}$.
This plot has considerable uncertainty due mainly to the large errors in the rate.
Here we suggest that an appropriate cadence for secure detection
of NS-mergers, and reasonable estimate of the merger time, is of the order of
1--2\,hr cadence.
The expected performances of some future surveys like the
Zwicky Transient Facility (ZTF), LSST, and {\it ULTRASAT}, are shown,
based on the following assumptions.
For ZTF we assumed the survey will observe about 1600\,deg$^{2}$
with weather efficiency of 0.8 (effectively reducing the field of view),
and limiting magnitude (averaged over all lunations) of about 20.1.
For LSST we assumed a dedicated survey for NS-merger events
with 5\% of the telescope time, observing about 100 fields (i.e., 960\,deg$^{2}$),
with 1\,hr cadence, weather efficiency of 0.8 and limiting magnitude of 24.
For {\it ULTRASAT} two options were considered.
The first is observing a single field continuously and coadding images on 3\,hr timescale
(ULTRASAT 3\,hr).
The second option is observing 6 fields, in rotation, with 1.5\,hr cadence
(ULTRASAT 6 fields).
We estimate that {\it ULTRASAT} may find of the order of 10 merger events per year.

The figure also shows the A-LIGO 8-$\sigma$ detection horizon (corresponding to 200\,Mpc),
and it's 6-$\sigma$ detection horizon (corresponding to 270\,Mpc;
the strain signal-to-noise ratio decreases linearly with the luminosity distance).
This suggests that future surveys like {\it ULTRASAT} and LSST hold the potential
for finding non triggered NS-merger events, and to trigger a LIGO search.
Specifically, given the known merger time and sky location, one can
search for 6-$\sigma$ events in the A-LIGO event streams,
and confirm them retroactively.
Such a strategy can significantly increase the number of future NS merger detections.

\section*{Acknowledgements} We thank Boaz Katz, Nahliel Wygoda, Yuri Levin \& Matias Zaldarriaga for insightful discussions during the conduction of this work, and Dan Kasen, Jennifer Barnes, Meng-Wu Ru, Dan Maoz, Anders Jerkstrand, Masaru Shibata \& Stephen Smartt for useful discussions following constructive comments on the draft. We thank Elena Pian for sharing the X-Shooter spectra. E.W. is partially supported by IMOS, ISF, and Minerva grants. E.O.O. is grateful for support by grants from the Willner Family Leadership Institute Ilan Gluzman (Secaucus NJ), ISF, Minerva, BSF, and BSF-transformative. A.G.-Y. is partially supported by the EU via ERC grant No. 725161, the ISF, the BSF Transformative program and by a Kimmel award. E.W., E.O.O and A.G.-Y. are partially supported by the I-Core program of the ICPB \& the ISF.

\onecolumn
\appendix
\section{A derivation of the model predictions}
\label{sec:ModelDer}

We denote the location of the diffusion-sphere, beyond which radiation escapes on a dynamical time scale, $\tau=c/v$, as $\tilde{m}_{\rm D}(t)=m_{\rm D}(t)M$. The velocity and radius of the mass element at which the diffusion-sphere lies are given by
\begin{equation}\label{eq:vD}
    v_{\rm D}(t)=v[m_{\rm D}(t)],\quad {\rm and}\quad r_{\rm D}(t)=v_{\rm D}(t)t.
\end{equation}
Similarly, we denote the location of the photosphere, $\tau=1$, by $\tilde{m}_{\rm ph}(t)=m_{\rm ph}(t)M$. The velocity and radius of the mass element at which the photosphere lies are given by
\begin{equation}\label{eq:vph}
    v_{\rm ph}(t)=v[m_{\rm ph}(t)],\quad {\rm and}\quad r_{\rm ph}(t)=v_{\rm ph}(t)t.
\end{equation}
Note that $v_{\rm ph}(t)$ is the velocity of the mass fraction at which the photosphere
lies at time $t$ (rather than the time derivative of $r_{\rm ph}$).
Using Eq.~(\ref{eq:tau}) and Eq.~(\ref{eq:tm}) we find
\begin{equation}\label{eq:m_diff}
  m_D(t)=\left(\frac{t}{t_{\rm M}}\right)^{\frac{2-\gamma}{1+\alpha}},\quad
  \frac{v_D(t)}{v_{\rm M}}=\left(\frac{t}{t_{\rm M}}\right)^{-\frac{(2-\gamma)\alpha}{1+\alpha}},\quad
  \frac{r_D(t)}{v_{\rm M}t_{\rm M}}=\left(\frac{t}{t_{\rm M}}\right)^{\frac{1-(1-\gamma)\alpha}{1+\alpha}}
\end{equation}
(for $t<t_{\rm M}$), and
\begin{equation}\label{eq:m_phot}
	m_{\rm ph} = \left(\frac{v_{\rm M}}{c}\right)^{\frac{1}{2\alpha+1}}
    \left(\frac{t}{t_{\rm M}}\right)^{\frac{2-\gamma}{2\alpha+1}},\quad
    	\frac{v_{\rm ph}(t)}{v_{\rm M}} = \left(\frac{v_{\rm M}}{c}\right)^{-\frac{\alpha}{2\alpha+1}} \left(\frac{t}{t_{\rm M}}\right)^{\frac{-\alpha(2-\gamma)}{2\alpha+1}},\quad
  \frac{r_{\rm ph}(t)}{ct_{\rm M}}=\left(\frac{v_{\rm M}}{c}\right)^{\frac{1+\alpha}{1+2\alpha}}
  \left(\frac{t}{t_{\rm M}}\right)^{\frac{1+\gamma\alpha}{1+2\alpha}}
\end{equation}
(for $t<t_{\rm ph}$). It is useful to note that for any mass fraction $m$, the time $t_m$ at which the diffusion sphere reaches $m$ is given by
\begin{equation}\label{eq:tm}
  t_m^2\equiv \frac{1}{4\pi(1+2\alpha)}\frac{\kappa(t=t_m) mM}{c v(m)}.
\end{equation}

We discuss first, in \S~\ref{sec:no_escape}, the emission of radiation for the case where the energy deposition of electrons and positrons is complete, $f_{ed}=1$. We then discuss the more general case of partial energy deposition in \S~\ref{sec:escape}.

\subsection{Complete e$^\pm$ local energy deposition}
\label{sec:no_escape}

Let us consider first times $t<t_{\rm M}$. Assuming that the initial radius was close to 10~km, the initial energy is completely lost to adiabatic expansion, and the energy of the ejecta is dominated by radioactive decay. At $r<r_{\rm D}$ ($m>m_{\rm D}$), we may neglect the diffusion (see below) and the energy per unit mass is governed by
\begin{equation}\label{eq:dEdt}
  \frac{d\epsilon}{dt}=-\frac{\epsilon}{t}+\dot{\varepsilon}_d.
\end{equation}
Here $\epsilon$ is the internal energy of the plasma. The first term in the energy equation represents adiabatic losses (assuming radiation domination). Noting that $t^{-1}d(\epsilon t)/dt =\epsilon/t+d\epsilon/dt$, we may integrate to find
\begin{equation}
	\epsilon(t) t- \epsilon(t_0) t_0 = \int_{t_0}^t dt' \dot{\varepsilon}_dt'.
\label{eq:Eintegral}
\end{equation}
For a power-law energy deposition rate and $t\gg t_0$ we have
\begin{equation}\label{eq:integral_power}
  \epsilon=
  \frac{1}{2-\beta}(1-f_{\nu\gamma})\dot{\varepsilon}_{\rm M} t_{\rm M}\left(\frac{t}{t_{\rm M}}\right)^{1-\beta}.
\end{equation}
The diffusion term,
\begin{equation}\label{eq:diff}
  \frac{d\epsilon}{dt}|_{\rm diff}=\frac{1}{4\pi r^2\rho}\partial_r\left[4\pi r^2 \frac{ c}{3\kappa\rho}\partial_r (\epsilon\rho)\right]
  \sim\frac{c\epsilon}{\kappa\rho r^2}\sim\frac{\epsilon}{\tau r/c}=\frac{1}{\tau v/c}\frac{\epsilon}{t},
\end{equation}
may be neglected for $(\tau v/c)\gg1$.

At $t<t_{\rm M}$, the luminosity is given by
\begin{equation}\label{eq:Lthick}
  L \approx M\epsilon\frac{dm_D}{dt}+Mm_D\dot{\varepsilon}.
\end{equation}
The first term represents the escape of photons from the hot ejecta due to the penetration of the "diffusion wave", while the second represents the energy deposited outside the diffusion sphere, $m<m_D$, and radiated on a dynamical time scale. At $t>t_{\rm M}$, after the diffusion sphere crosses through most of the mass of the ejecta and $m_{\rm D}=1$, the luminosity is given by the second term, $L\approx M\dot{\varepsilon}$. We therefore find
\begin{eqnarray}\label{eq:LM}
  L &\approx& M\dot{\varepsilon}_{\rm M}\left(\frac{t}{t_{\rm M}}\right)^{-\beta}\left\{\begin{array}{ll}
                      \left[1+\frac{2-\gamma}{(1+\alpha)(2-\beta)}\right]
                      (t/t_{\rm M})^{(2-\gamma)/(1+\alpha)}, & \hbox{$ t<t_{\rm M}$;} \\
                      1, & \hbox{$t_{\rm M}<t$.}
                    \end{array}
                             \right.
\end{eqnarray}

\subsection{Partial e$^\pm$ local energy deposition}
\label{sec:escape}

The mass fraction $m_e$ at which $f_{ed}=1$ is approximately given by
\begin{equation}\label{eq:m_gamma}
  m_e^{1+3\alpha}=\left(\frac{t}{t_\varepsilon}\right)^2=X_{\rm M}\left(\frac{t}{t_{\rm M}}\right)^2,
\end{equation}
and
\begin{equation}\label{eq:m_Dgamma}
  \left(\frac{m_D}{m_e}\right)^{1+3\alpha}
  =X_{\rm M}^{-1}\left(\frac{t}{t_{\rm M}}\right)^s,\quad s=\frac{4\alpha-(1+3\alpha)\gamma}{(1+\alpha)}.
\end{equation}
The time $t_e$, beyond which $f_{ed}<1$ for mass fraction $m$, is
\begin{equation}\label{eq:te}
  \frac{t_e}{t_\varepsilon}=m^{(1+3\alpha)/2}.
\end{equation}
At $t>t_e$ the energy deposition at $m$ is suppressed by $(t/t_e)^{-2}$.
The time $t_{eD}$ at which $m_e=m_D$ is
\begin{equation}\label{eq:t_e_D}
  \frac{t_{e D}}{t_{\rm M}}=X_{\rm M}^{1/s}.
\end{equation}

Four different cases should be considered:
\begin{enumerate}
  \item $s>0$, $t_{\rm M}<t_\varepsilon$: $t_{e D}<t_{\rm M}<t_\varepsilon$, for which $m_D<m_e$ at $t<t_{e D}$ and $m_D>m_e$ at $t>t_{e D}$;
  \item $s<0$, $t_{\rm M}<t_\varepsilon$: $t_{\rm M} <t_\varepsilon$, for which $m_D\ge m_e$;
  \item $s>0$, $t_{\rm M}>t_\varepsilon$: $t_\varepsilon<t_{\rm M}$, for which $m_D<m_e$;
  \item $s<0$, $t_{\rm M}>t_\varepsilon$: $t_{e D}<t_\varepsilon<t_{\rm M}$, for which $m_D>m_e$ at $t<t_{e D}$ and $m_D\le m_e$ at $t>t_{e D}$.
\end{enumerate}

Let us consider first the behavior at $t>t_{\rm M}$. At this stage $L$ is given by (see \S~\ref{sec:no_escape})
\begin{equation}\label{eq:Ld}
  L=M\dot{\varepsilon}_d,
\end{equation}
where the deposited energy is
\begin{equation}\label{eq:Ed}
  \dot{\varepsilon}_d=
  (1-f_{\nu\gamma})\dot{\varepsilon}\min\left[1,\left(\frac{t}{t_\varepsilon}\right)^{-2}\right].
\end{equation}
This yields
\begin{equation}\label{eq:L_late}
  L=M(1-f_{\nu\gamma})\dot{\varepsilon}_{\rm M}\left(\frac{t}{t_{\rm M}
  }\right)^{-\beta}
  \min\left[1,\left(\frac{t}{t_\varepsilon}\right)^{-2}\right].
\end{equation}

Let us consider next the regime $t_\varepsilon<t<t_{\rm M}$. In this regime the diffusion sphere did not yet cross the ejecta, hence $L\approx M\dot{m}_D\varepsilon(m_D)$, and radiation is emitted at time $t$ from a mass shell $m_D(t)$ for which $f_{ed}$ became smaller than unity at $t_e<t$. The internal energy is given in this case by
\begin{equation}\label{eq:Ethin}
  \epsilon(m,t)\approx (1-f_{\nu\gamma})\dot{\varepsilon}_{\rm M}t_{\rm M}\left(\frac{t_e}{t_{\rm M}}\right)^{1-\beta}
  \left(\frac{t}{t_e}\right)^{-1}
  =(1-f_{\nu\gamma})\dot{\varepsilon}_{\rm M}t_{\rm M}
  \left(\frac{t_e}{t_\varepsilon}\frac{t_\varepsilon}{t_{\rm M}}\right)^{2-\beta}
  \left(\frac{t}{t_{\rm M}}\right)^{-1}
\end{equation}
(the internal energy evolves like $t^{1-\beta}$ at $t<t_e$, and since the deposition drops faster than $1/t^2$ at $t>t_e$, its later evolution follows adiabatic expansion -- $\epsilon\propto 1/t$; we neglect here the effects of diffusion at $r<r_D$, as may be justified in a manner similar to that of the preceding section). Using $t_e/t_\varepsilon=m^{(1+3\alpha)/2}$ and $m_D=(t/t_{\rm M})^{(2-\gamma)/(1+\alpha)}$ we obtain
\begin{equation}\label{eq:L_interm}
  L\approx\frac{2-\gamma}{1+\alpha}M(1-f_{\nu\gamma})\dot{\varepsilon}_{\rm M}
  \left(\frac{t_\varepsilon}{t_{\rm M}}\right)^{2-\beta}
  \left(\frac{t}{t_{\rm M}}\right)^{\frac{2-\gamma}{1+\alpha}-\beta+s(1-\frac{\beta}{2})}.
\end{equation}
Note that for $t_\varepsilon<t_{\rm M}$, Eq.~(\ref{eq:L_interm}) gives at $t=t_{\rm M}$ a luminosity larger by a factor $(t_{\rm M}/t_\varepsilon)^\beta$ than that give by Eq.~(\ref{eq:L_late}). This reflects the fact that in this case the energy deposition rate in the ejecta at $t_{\rm M}$, $\dot{\varepsilon}_d(t_{\rm M})=\dot{\varepsilon}_{\rm M}(t_{\rm M}/t_\varepsilon)^{-2}$, while the ejecta energy density is determined by the (higher) deposition rate at $t=t_\varepsilon<t_{\rm M}$, $\varepsilon(t_{\rm M})/t_{\rm M}=\dot{\varepsilon}_{\rm M}(t_{\rm M}/t_\varepsilon)^{\beta-2}$. As the photons escape the ejecta at $t=t_{\rm M}$, there is a drop in $L$ in the transition to the $t>t_{\rm M}$ regime.

Let us consider next $t<\min[t_{\rm M},t_\varepsilon]$. For $t_{\rm M}>t_\varepsilon$, $m_D<m_e$ at all time for $s>0$ and at $t>t_{e D}$ for $s<0$. For $m_D<m_e$, radiation is emitted from regions for which $f_{ed}$ dropped below unity at earlier time, hence the energy density is given by Eq.~(\ref{eq:Ethin}) and $L$ by Eq.~(\ref{eq:L_interm}). For $t_{\rm M}<t_\varepsilon$, $m_D>m_e$ at all time for $s<0$ and at $t>t_{e D}$ for $s>0$. For $m_D>m_e$ radiation is emitted from regions where $f_{ed}=1$, hence the energy density is
\begin{equation}\label{eq:Ethick}
  \varepsilon(m,t)\approx(1-f_{\nu\gamma}) \dot{\varepsilon}_{\rm M}t_{\rm M}\left(\frac{t}{t_{\rm M}}\right)^{1-\beta}
\end{equation}
and
\begin{equation}\label{eq:L_thick}
  L\approx M\dot{m_D}\varepsilon(m_D)=\frac{2-\gamma}{1+\alpha}M(1-f_{\nu\gamma})\dot{\varepsilon}_{\rm M}
  \left(\frac{t}{t_{\rm M}}\right)^{\frac{2-\gamma}{1+\alpha}-\beta}.
\end{equation}

Summarizing our results we have for $t_{\rm M}<t_\varepsilon$ ($X_{\rm M}<1$)
\begin{equation}\label{eq:L_tM_tE}
  L=M(1-f_{\nu\gamma})\dot{\varepsilon}_{\rm M}
  \left\{
    \begin{array}{ll}
      \eta X_{\rm M}^{-1+\beta/2}(t/t_{\rm M})^{\frac{2-\gamma}{1+\alpha}-\beta+s(1-\frac{\beta}{2})}, &
             \hbox{$t<t_{e D}=X_{\rm M}^{1/s}t_{\rm M}$ and $s>0$;} \\
      \eta(t/t_{\rm M})^{\frac{2-\gamma}{1+\alpha}-\beta}, &
             \hbox{$t<t_{\rm M}$ (and $t_{e D}<t$ for $s>0$);} \\
      (t/t_{\rm M})^{-\beta}, & \hbox{$t_{\rm M}<t<t_\varepsilon=X_{\rm M}^{-1/2}t_{\rm M}$;} \\
      X_{\rm M}^{\beta/2}(t/t_\varepsilon)^{-\beta-2}, & \hbox{$t_\varepsilon<t$,}
    \end{array}
  \right.
\end{equation}
and for $t_{\rm M}>t_\varepsilon$ ($X_{\rm M}>1$)
\begin{equation}\label{eq:L_tE_tM}
  L=M(1-f_{\nu\gamma})\dot{\varepsilon}_{\rm M}
  \left\{
    \begin{array}{ll}
      \eta(t/t_{\rm M})^{\frac{2-\gamma}{1+\alpha}-\beta}, &
             \hbox{$t<t_{e D}=X_{\rm M}^{1/s}t_{\rm M}$ and $s<0$;} \\
      \eta X_{\rm M}^{-1+\beta/2}(t/t_{\rm M})^{\frac{2-\gamma}{1+\alpha}-\beta+s(1-\frac{\beta}{2})}, &
        \hbox{$t<t_{\rm M}$ (and $t_{e D}<t$ for $s<0$);} \\
      X_{\rm M}^{-1}(t/t_{\rm M})^{-\beta-2}, & \hbox{$t_{\rm M}<t$.}
    \end{array}
  \right.
\end{equation}
Here $\eta\equiv1+\frac{2-\gamma}{(1+\alpha)(2-\beta)}$ (see preceding section, Eq.~\ref{eq:LM}).

\section{Detailed description of the data analysis results}
\label{sec:data}

\begin{figure}
\centerline{\includegraphics[width=8cm]{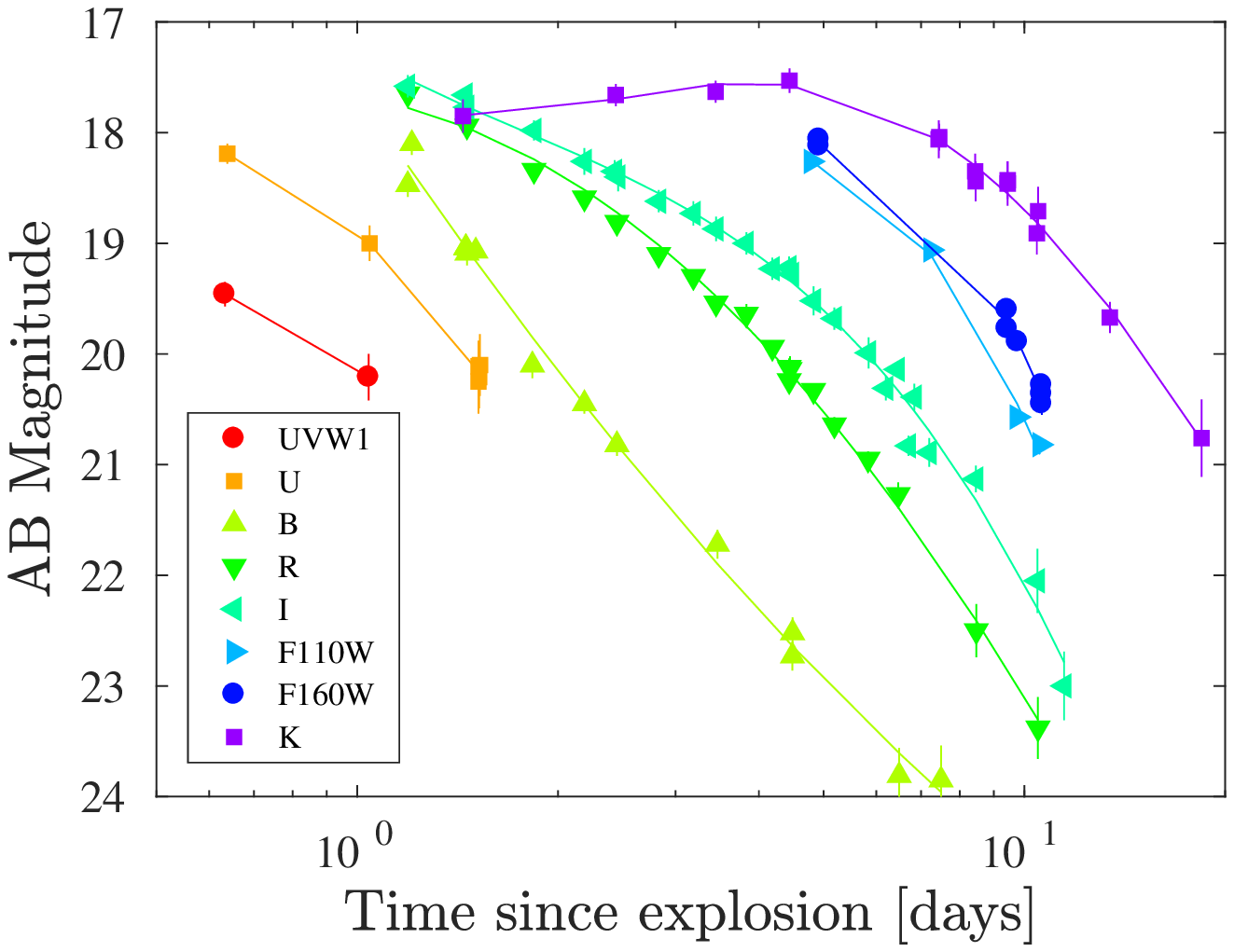}}
\caption{Light curves of GW\,170817 in selected bands and their polynomial fits (see also Table~\ref{tab:MultiBand}).
\label{fig:LC_Bands}}
\end{figure}

Tables \ref{tab:MultiBand}-\ref{tab:BolLumSpec} and Fig.~\ref{fig:LC_Bands} provide the detailed results of the data analysis described in \S~\ref{sec:obs}.

\begin{deluxetable}{lrlll}
\tablecolumns{5}
\tablewidth{0pt}
\tablecaption{Polynomial fit to multi-band data}
\tablehead{
\colhead{Band}             &
\colhead{$\min{t}$}        &
\colhead{$\max{t}$}        &
\colhead{rms}              &
\colhead{Poly. Coef.}      \\
\colhead{}                 &
\colhead{day}              &
\colhead{day}              &
\colhead{}                 &
\colhead{}                 \\
}
\startdata
B         &    0.990 &  7.690 & 0.14 &     $-2.010$, 8.994, 17.627 \\
F110W     &    4.590 & 10.690 & 0.09 &     18.302, $-23.438$, 25.726 \\
F160W     &    4.723 & 10.723 & 0.07 &     22.117, $-31.192$, 29.084 \\
F336W     &    5.100 &  5.500 & 0.10 &     25.050 \\
F475W     &    7.916 & 11.516 & 0.04 &     $-59.078$, 122.087, $-39.029$ \\
F606W     &   11.200 & 11.600 & 0.10 &     23.090 \\
F814W     &   11.200 & 11.600 & 0.10 &     22.320 \\
H         &    0.361 & 14.561 & 0.17 &      3.014, $-1.848$, 1.668, $-0.806$, 17.692 \\
I         &    1.090 & 11.490 & 0.13 &      5.112, $-3.814$, 3.508, 17.270 \\
J         &    0.471 & 10.471 & 0.10 &      3.657, 7.704, $-16.040$, 5.934, 3.008, $-1.034$, 17.584 \\
K         &    1.340 & 18.440 & 0.08 &     $-4.715$, 17.228, $-15.540$, 4.224, 17.499 \\
Ks        &    0.370 & 25.470 & 0.17 &     $-0.256$, 3.303, $-0.687$, $-1.764$, 18.136 \\
M2        &    0.527 &  1.127 & 0.00 &      6.088, 22.352 \\
R         &    0.990 & 10.590 & 0.08 &      4.495, 0.926, 17.680 \\
U         &    0.638 &  1.638 & 0.06 &      8.744, 5.347, 18.900 \\
V         &    0.471 & 10.671 & 0.10 &     $-3.711$, 10.006, $-8.364$, $-1.384$, 6.864, 2.647, 17.571 \\
W1        &    0.433 &  1.233 & 0.20 &      3.480, 20.150 \\
W2        &    0.600 &  0.700 & 0.25 &     21.130 \\
Y         &    0.480 & 11.580 & 0.26 &      3.998, $-0.247$, $-0.730$, 17.610 \\
g         &    0.477 &  7.577 & 0.14 &     11.766, $-13.658$, 9.020, $-12.778$, 8.144, 4.902, 17.601 \\
i         &    0.452 & 10.652 & 0.17 &      1.023, 1.930, 1.302, 17.527 \\
r         &    0.464 & 12.564 & 0.20 &     $-2.008$, 4.534, 2.255, 17.529 \\
u         &    1.451 &  3.651 & 0.30 &    $-12.812$, 17.255, 17.484 \\
y         &    0.702 &  6.802 & 0.05 &      2.143, 0.504, 17.399 \\
z         &    0.471 & 14.571 & 0.27 &      3.444, $-0.354$, 0.609, 17.622 \\
\enddata
\tablecomments{Polynomial fits to the multi-band light curves of AT\,2017gfo
in $\log{t}$-magnitude space.
$\min{t}$ and $\max{t}$ are the minimum and maximum times (days since merger)
for which the polynomial fitting is relevant.
rms is the best fit rms value in magnitudes,
while polynomial coef. corresponds to the high to low polynomila order.}
\label{tab:MultiBand}
\end{deluxetable}

\begin{deluxetable}{lrlllllllll}
\tablecolumns{5}
\tablewidth{0pt}
\tablecaption{Bolometric luminosity of AT\,2017gfo}
\tablehead{
\colhead{t}                &
\colhead{$L_{\rm int}$}    &
\colhead{$L_{\rm bb}$}     &
\colhead{$-\Delta{L}$}     &
\colhead{$+\Delta{L}$}     &
\colhead{$T_{\rm eff}$}    &
\colhead{$-\Delta{T}$}     &
\colhead{$+\Delta{T}$}     &
\colhead{$R_{\rm phot}$}   &
\colhead{$\chi^{2}$/dof}    &
\colhead{$\min\lambda--\max\lambda$}   \\
\colhead{day}              &
\colhead{erg\,s$^{-1}$}    &
\colhead{erg\,s$^{-1}$}    &
\colhead{erg\,s$^{-1}$}    &
\colhead{erg\,s$^{-1}$}    &
\colhead{K}    &
\colhead{K}    &
\colhead{K}    &
\colhead{cm}   &
\colhead{}       &
\colhead{\AA}
}
\startdata
 0.5 &  $7.82\times10^{41}$ & $1.2\times10^{42}$ & $9.6\times10^{40}$ & $2.9\times10^{40}$ & 10266 &  367 &  304 &  $3.7_{-0.1}^{+0.2}\times10^{14}$ &     86.5/7 &  2613 -- 21908\\
 0.6 &  $8.45\times10^{41}$ & $1.3\times10^{42}$ & $1.4\times10^{41}$ & $2.0\times10^{41}$ & 10809 &  659 &  764 &  $3.6_{-0.3}^{+0.3}\times10^{14}$ &    311/7 &  2079 -- 21908\\
 0.8 &  $7.31\times10^{41}$ & $7.0\times10^{41}$ & $3.4\times10^{40}$ & $1.8\times10^{40}$ &  7093 &  142 &  146 &  $6.0_{-0.2}^{+0.2}\times10^{14}$ &   2022/9 &  2254 -- 21908\\
 1.0 &  $6.15\times10^{41}$ & $5.5\times10^{41}$ & $6.8\times10^{39}$ & $2.8\times10^{40}$ &  6351 &  115 &  123 &  $6.7_{-0.3}^{+0.3}\times10^{14}$ &   2721/11 &  2254 -- 21908\\
 1.2 &  $4.97\times10^{41}$ & $4.8\times10^{41}$ & $2.3\times10^{40}$ & $5.8\times10^{37}$ &  5707 &   73 &   81 &  $7.6_{-0.3}^{+0.2}\times10^{14}$ &   1358/12 &  2613 -- 21908\\
 1.5 &  $3.69\times10^{41}$ & $3.6\times10^{41}$ & $2.6\times10^{37}$ & $1.8\times10^{40}$ &  4967 &   39 &   39 &  $8.9_{-0.1}^{+0.3}\times10^{14}$ &   1434/14 &  3560 -- 21908\\
 2.5 &  $2.44\times10^{41}$ & $2.3\times10^{41}$ & $5.5\times10^{39}$ & $8.5\times10^{39}$ &  3751 &   86 &   92 &  $1.2_{-0.1}^{+0.1}\times10^{15}$ &   2821/13 &  3560 --  21908\\
 3.5 &  $1.87\times10^{41}$ & $1.7\times10^{41}$ & $8.3\times10^{39}$ & $6.5\times10^{39}$ &  3160 &   86 &   90 &  $1.5_{-0.1}^{+0.1}\times10^{15}$ &   5490/13 &  3560 -- 21908\\
 4.5 &  $1.46\times10^{41}$ & $1.4\times10^{41}$ & $8.1\times10^{39}$ & $8.6\times10^{39}$ &  2836 &   84 &   86 &  $1.7_{-0.1}^{+0.1}\times10^{15}$ &   7513/12 &  4473 --  21908\\
 5.5 &  $1.16\times10^{41}$ & $1.0\times10^{41}$ & $8.6\times10^{39}$ & $9.4\times10^{39}$ &  2951 &  101 &  107 &  $1.3_{-0.1}^{+0.1}\times10^{15}$ &  32631/15 &  3351 --  21908\\
 6.5 &  $8.84\times10^{40}$ & $9.1\times10^{40}$ & $5.5\times10^{39}$ & $5.8\times10^{39}$ &  2505 &   64 &   66 &  $1.7_{-0.1}^{+0.1}\times10^{15}$ &  10884/14 &  4473  --  21908\\
 7.5 &  $6.53\times10^{40}$ & $7.0\times10^{40}$ & $6.6\times10^{39}$ & $6.4\times10^{39}$ &  2452 &   71 &   73 &  $1.6_{-0.1}^{+0.2}\times10^{15}$ &  13313/13 &  4473  --  21908\\
 8.5 &  $4.69\times10^{40}$ & $4.5\times10^{40}$ & $6.7\times10^{39}$ & $8.5\times10^{39}$ &  2848 &   96 &  102 &  $9.5_{-1.3}^{+1.5}\times10^{14}$ &  53817/12 &  4755  --  21908\\
 9.5 &  $3.36\times10^{40}$ & $2.9\times10^{40}$ & $4.9\times10^{39}$ & $6.4\times10^{39}$ &  2857 &  106 &  114 &  $7.6_{-1.1}^{+1.3}\times10^{14}$ &  66549/12 &  4755  --  21908\\
10.5 &  $2.47\times10^{40}$ & $2.0\times10^{40}$ & $4.2\times10^{39}$ & $5.0\times10^{39}$ &  2963 &  137 &  153 &  $5.8_{-1.1}^{+1.3}\times10^{14}$ &  88875/11 &  4755  --  21908\\
11.5 &  $1.96\times10^{40}$ & $3.0\times10^{40}$ & $7.5\times10^{39}$ & $1.0\times10^{40}$ &  2808 &  131 &  146 &  $8.0_{-1.7}^{+2.1}\times10^{14}$ &  29503/7 &  4755  --  21908\\
12.5 &  $1.40\times10^{40}$ & $3.0\times10^{40}$ & $5.0\times10^{39}$ & $6.9\times10^{39}$ &  2426 &  308 &  418 &  $1.1_{-0.2}^{+0.3}\times10^{15}$ &   2163/3 &  6183  --  21908\\
13.5 &  $9.64\times10^{39}$ & $2.1\times10^{40}$ & $3.1\times10^{39}$ & $4.9\times10^{39}$ &  1641 &  329 &  576 &  $1.9_{-0.8}^{+1.3}\times10^{15}$ &   1027/2 &  8922  --  21908\\
14.5 &  $7.33\times10^{39}$ & $1.7\times10^{40}$ & $2.8\times10^{39}$ & $4.9\times10^{39}$ &  1625 &  353 &  675 &  $1.7_{-0.8}^{+1.3}\times10^{15}$ &   1311/2 &  8922  --  21908\\
15.5 &  $3.28\times10^{39}$ & $7.7\times10^{38}$ & $7.2\times10^{38}$ & $7.0\times10^{40}$ &  1917 &  755 & 3560 &  $1.9_{-0.8}^{+3.9}\times10^{14}$ &      0.4/0 & 21908  --  21908\\
16.5 &  $2.61\times10^{39}$ & $7.7\times10^{38}$ & $7.2\times10^{38}$ & $7.03\times10^{40}$ &  1917 &  755 & 3560 &  $1.9_{-0.8}^{+3.9}\times10^{14}$&      0.9/0 & 21908  --  21908\\
\enddata
\tablecomments{Bolometric luminosity of AT\,2017gfo as a function of time.
$L_{\rm int}$ is the bolometric luminosity estimated using trapezoidal integration of the photometric data;
$L_{\rm bb}$ is the bolometric luminosity estimated using a blackbody fit to the photometric observations,
while $T$ and $R$ are the effective tempearture and photospheric radius.
$-\Delta$ and $+\Delta$ corresponds to lower and upper 1-$\sigma$ uncertanty estimated using margenlizing over all
other parameters.
$\chi^{2}$ and dof corresponds to the blackbody fit.
$\min\lambda$ and $\max\lambda$ are the minimum and maximum central wavelength of the band used in the estimation. The errors do not include any systematic uncertainty in the photometric calibration
(presumably better than 10\%).}
\label{tab:BolLum}
\end{deluxetable}

\begin{deluxetable}{llll}
\tablecolumns{4}
\tablewidth{0pt}
\tablecaption{Spectra blackbody fit}
\tablehead{
\colhead{t}                &
\colhead{$T$}              &
\colhead{$L_{0.3-2.4\mu{\rm m}}$} &
\colhead{IR fraction}      \\
\colhead{day}              &
\colhead{K}                &
\colhead{erg\,s$^{-1}$}    &
\colhead{}
}
\startdata
  1.441  &  4996 &  $2.88\times10^{41}$ & 0.37\\
  3.451  &  3059 &  $1.44\times10^{41}$ & 0.68\\
  5.451  &  2646 &  $1.02\times10^{41}$ & 0.75\\
  6.451  &  2624 &  $7.70\times10^{40}$ & 0.75\\
  7.451  &  2589 &  $5.56\times10^{40}$ & 0.75\\
  8.451  &  2654 &  $3.71\times10^{40}$ & 0.73\\
  9.441  &  2703 &  $2.48\times10^{40}$ & 0.73\\
 10.441  &  2975 &  $1.49\times10^{40}$ & 0.70\\
\enddata
\tablecomments{Properties of the spectra of AT\,2017gfo.
All spectra were obtained using X-Shooter and are
from \cite{2017Natur.551...67P} and
\cite{2017Natur.551...75S}.
$T$ is the best fit blackbody temperature, $L$ is the integrated luminosity,
and IR fraction is the fraction of the luminosity in the IR $>1\mu$m to the
integrated luminosity.}
\label{tab:BolLumSpec}
\end{deluxetable}

\twocolumn
\bibliographystyle{mnras}

\bsp	
\label{lastpage}
\end{document}